\documentstyle[amscd]{amsart}

\def\A{{\Bbb A}}
\def\N{{\Bbb N}}
\def\Z{{\Bbb Z}}
\def\Q{{\Bbb Q}}
\def\R{{\Bbb R}}
\def\C{{\Bbb C}}
\def\P{{\Bbb P}}

\def\s{\section}
\def\ss{\subsection}
\def\st{\subsection}
\def\sss{\subsubsection}
\def\SC{Section }
\def\GS{section }

\newtheorem{theorem}{Theorem}[section]
\newtheorem{lemma}[theorem]{Lemma}
\newtheorem{proposition}[theorem]{Proposition}
\newtheorem{corollary}[theorem]{Corollary}
\newtheorem{definition}[theorem]{Definition}

\newtheorem{conjecture}[theorem]{Conjecture}

\newtheorem{remark}[theorem]{Remark}

\theoremstyle{definition}
\theoremstyle{remark}

\begin{document}

\title{Vector Braids}

\author{Vincent~L.~Moulton}

\address{Department of Mathematics\\
Duke University\\ Durham, NC 27708-0320}

\thanks{}

\email{moulton@@math.duke.edu}
\date{July 15th, 1994}

\maketitle

\begin{abstract}
In this paper we define a new family of groups which generalize
the {\it classical  braid groups on} $\C $. We
denote this family by $\{B_n^m\}_{n \ge m+1}$ where $n,m \in \N$.
The family $\{ B_n^1 \}_{n \in \N}$ is the set of classical braid groups on $n$
strings.
The group $B_n^m$ is the set of motions of $n$ unordered points in $\C^m$,
so that at any time during the motion, each $m+1$ of the points
span the whole of $\C^m$ as an affine space.
There is a  map from $B_n^m$
to the symmetric group on $n$ letters. We
let $P_n^m$ denote the kernel of this map.
In this paper we are mainly interested
in understanding $P_n^2$. We give a presentation of a group $PL_n$ which maps
surjectively onto $P_n^2$. We also show the surjection
$PL_n \to P_n^2$ induces an isomorphism on first and
second integral homology  and conjecture
that it is an isomorphism.
We then find an infinitesimal presentation of the group $P_n^2$.
Finally, we also consider the analagous groups where
points lie in $\P^m$ instead of $\C^m$. These groups generalize of the
classical
braid groups on the sphere.
\end{abstract}

\s{Introduction}\label{intro}

Let $\A^m$ denote $m$-dimensional, complex affine space.
Let $X_n^m$ be the space of ordered $n$-tuples of elements of $\A^m$, with $n
\ge m+1$
so that each $m+1$ of the components of
each $n$-tuple span the whole of  $\A^m$ in the sense
of affine geometry. The symmetric group on $n$ letters,
$\Sigma_n$, acts on $X_n^m$ via permuting the components of each point.
This action is fixed point free. We can therefore form the
quotient space $X_n^m/\Sigma_n$.

\begin{definition}
Let $B_n^m = \pi_1(X_n^m/\Sigma_n)$ and $P_n^m = \pi_1(X_n^m)$.
Call these groups {\it the group of n stringed vector braids on $\A^m$}
and {\it the group of n stringed pure vector braids on $\A^m$}, respectively.
\end{definition}

The long exact sequence of a fibration gives us the short exact sequence
\begin{equation} \label{ses}
1 \to P_n^m \to B_n^m \to \Sigma_n \to 1.
\end{equation}

The space $X_n^1$ is the well
known {\it configuration space} of $n$ points in $\C$.
In general we can describe the space $X_n^m$ as
$$
X_n^m = \underbrace{\A^m \times \dots \times \A^m}_{n} - \Delta
$$
where
$$
\Delta = \{ (x_1, \dots , x_n ) \, | \, x_i \in \A^m \, ,\, \text{span}
\{x_{i_1},\dots , x_{i_{m+1}}\} \neq \A^m \}.
$$
In the case $m=1$, the set $\Delta$ is simply the ``fat diagonal''. In general
we will call the
set $\Delta$ the {\it dependency locus}. Choose coordinates $(x_1,\dots ,
x_n)$ for $\A^m$ i.e. an isomorphism between
$\A^m$ and $\C^m$.
Let $x_i = (z_{1i},\dots, z_{mi})$, where
$1 \le i \le n$ and $z_{ij} \in \C$.  Then the defining
equations  of $\Delta$
are all possible $(m+1) \times (m+1)$ minors
$$
\Delta_{i_1\dots i_{m+1}} =
\left|
\begin{array}{ccc}
 1         & \cdots   &        1           \\
 z_{1i_1}  & \cdots   &     z_{1i_{m+1}}   \\
\vdots     &          &     \vdots         \\
 z_{mi_1}  & \cdots   &     z_{mi_{m+1}}
\end{array}
\right|
$$
of the matrix
$$
\left[
\begin{array}{cccc}
 1         &  1         & \cdots   &        1     \\
 z_{11}    &  z_{12}    & \cdots   &     z_{1n}   \\
\vdots     &  \vdots    &          &     \vdots   \\
 z_{m1}    &  z_{m2}    & \cdots   &     z_{mn}
\end{array}
\right]
$$

\begin{remark} \begin{em}
We can also consider motions of $n$ points in $\P^m$ instead of $\A^m$ (see \SC
\ref{affine}).
\end{em} \end{remark}

The group $P_n^1$ is the classical {\it pure braid group of $n$ strings on
$\C$}.
The pure braid group has a very nice presentation, which may be understood
geometrically \cite{birman}.
The main aim of this paper is to discover
a geometrical presentation for the group $P_n^2$ analogous to that
of the classical pure braid group.
Elements of $P_n^2$ may be thought of as motions of $n$ points in $\A^2$
so that at any time during this motion no three of the $n$ points lie on a
line.

We can now informally state the main results of this paper.
We define a  group $PL_n$ via a presentation, and a surjective homomorphism
$\varphi_n :  PL_n \to P_n^2$.
The presentation of $PL_n$ is given in Definition \ref{pln}.
Theorems \ref{main theorem2} and \ref{main theorem3} state that
the homomorphism $\varphi_n$ induces isomorphisms on the first
and second integral homology groups\footnote{Tomohide Terasoma from Tokyo
Metropolitan University has recently found relations in
the fundamental group of the space of $n$-tuples of points
in $\P^m$, such that each  $m$ of the points span a hyperplane in $\P^m$
\cite{tt}.}.
In light of these facts the following conjecture seems reasonable.

\begin{conjecture} \label{hope}
The homomorphism $\varphi_n : PL_n \to P_n^2$ is an isomorphism.
\end{conjecture}

Despite the fact that it appears to be difficult to
prove Conjecture \ref{hope}, it is relatively straightforward to
find an infinitesimal presentation for $P_n^2$.
In Proposition \ref{provin}
we prove that the completion of the
group algebra $\C [P_n^2]$ with respect to
the augmentation ideal is isomorphic to the non-commutative
power series ring in the indeterminates $X_{ijk}$, $1 \le i < j <k \le n$,
modulo the
two  sided ideal generated by the relations
$$
[X_{ijk}, X_{jkl} + X_{ikl} + X_{ijl} ]  \quad i<j<k<l;
$$
$$
[X_{jkl}, X_{ijk} + X_{ikl} + X_{ijl} ]  \quad i<j<k<l;
$$
$$
[X_{ikl}, X_{ijk} + X_{ijl} + X_{jkl} ]  \quad i<j<k<l;
$$
$$
[X_{ijk}, X_{12k} + \dots + X_{k,n-1,n} ] \quad 1 \le k \le n;
$$
$$
[X_{ijk},X_{rst} ] \quad i,j,k,r,s,t \quad \text{distinct.}
$$

We now summarize the contents of this paper. First
it is useful to recall one method for finding a presentation of
the classical pure braid group, $P_n = P_n^1$.
Recall that the map
$$
p_n^1 : X_n^1 \to X_{n-1}^1
$$
\noindent which forgets the last point in each $n$-tuple is a fibration
with fiber equal to $\C$ less $n-1$ points \cite{F-N}.
For $n \ge 3$, the long exact sequence
for a fibration  provides  us with  the following sequence:
\begin{equation} \label{pureseq}
  1 \to L_{n-1} \to P_n \to P_{n-1} \to 1
\end{equation}
\noindent where $L_{n-1}$ is a free group on ${n-1}$ generators.
Let $a_{in}$, $ 1 \le i \le n-1$ denote the
loop $a_{in}$ loop in which the $n$th point goes
around the $i$th point in  the punctured $\C$. Then the set of
loops $ \{ a_{in} | 1 \le i \le n-1 \}$ generates $L_{n-1}$.
We picture the loop $a_{in}$ as an element of $P_n$ in Figure \ref{puregenpic}.

\begin{figure}[b]
\vskip 1in
\caption{The generator $a_{in}$ of $P_n$} \label{puregenpic}
\end{figure}

Using  sequence (\ref{pureseq}) and the fact that $P_2 = \Z$, we can
inductively
prove that the group $P_n$ admits a presentation with generators
$$
a_{ij}, \, 1 \le i< j \le n,
$$
and  defining relations,
\vskip .1in
%\begin{center}
$a_{ij}^{-1} \, a_{rs} \, a_{ij} =$  \hfill
%\end{center}
$$
\left\{  \begin{array}{ll}
a_{rs}  &      \mbox{if \ } r<i<j<s \text{ or } i<j<r<s; \\
a_{is}\,a_{rs}\,a_{is}^{-1} & \mbox{if \ } i<r=j<s;\\
a_{is}\,a_{js}\,a_{rs}\,a_{js}^{-1}\,a_{is}^{-1} & \mbox{if \ } r=i<j<s;\\
a_{is}\,a_{js}\,a_{is}^{-1}\,a_{js}^{-1}a_{rs}\,a_{js}
\,a_{is}\,a_{js}^{-1}\,a_{is}^{-1} & \mbox{if \ } i<r<j<s .
\end{array} \right. \hfill
$$
\vskip 0.05in
\noindent Since the generators of $P_{n-1}$ clearly lift
to $P_n$, finding a presentation for $P_n$ at each stage involves three main
operations.
First, we have to add the generators from the
fiber group $L_{n-1}$ to the group $P_n$.
Second, we have to add relations to $P_n$
obtained by conjugating each generator of $L_{n-1}$
by the generators of  $P_{n-1}$. Finally, we have to
lift  relations from $P_{n-1}$ to $P_n$.

The way in which we find presentations for
the groups $PL_n$ will be modelled on this approach, although, as we
shall see, there will be significant complications.

In \SC \ref{affine} we define the anologues of $P_n^m$ with $\A^m$ replaced by
$\P^m$,
and compare these groups to $P_n^m$ and $B_n^m$, respectively.
In \SC \ref
{action} we discuss the groups $P_n^m$ and $B_n^m$ for $m\le n+2$.
In \SC  \ref{forget} we show that the map
$$
p_n^2 : X_n^2 \to X_{n-1}^2
$$
\noindent obtained by forgetting the last point in each
$n$-tuple is not a fibration for $n \ge 5$. However, we will
be able to repair this defect in some sense by using the fact that these maps
are
fibrations except over a set of complex codimension one. A
consequence of this fact is Corollary \ref{exseq},
which states that there is  an exact sequence
$$
   \pi_1(GF_n^2)  \to  P_n^2 \to P_{n-1}^2 \to 1,
$$
\noindent where $GF_n^2$ denotes the generic fiber of the map $p_n^2$.

We use this sequence to find relations within the group $P_n^2$.
However, there are two major complications which did not occur
when we found  a presentation for the classical pure braid group.
First, the fiber is more intricate than in the braid case; it is the complement
of lines
in $\A^2$ rather than a punctured copy of $\C$.
Second, and more importantly, we do not know if
the group  $\pi_1(GF_n^2)$ injects into  $P_n^2$. If this were the
case, then Conjecture \ref{hope} would be true.

In \SC \ref{around} we give a way of describing loops in the
complexification of a real arrangement of lines in $\C^2$.
In \SC \ref{fiber} we use some techniques from stratified Morse theory
to find  nice presentations for  the fiber groups, $\pi_1(GF_n^2)$.
In \SC \ref{main} we define the group $PL_n$ and a surjective homomorphism
$\varphi_n : PL_n \to  P_n^2$. We then state the main theorems
of this paper, Theorems \ref{main theorem1}, \ref{main theorem2} and \ref{main
theorem3},
and prove Theorems  \ref{main theorem2} and \ref{main theorem3}.
In \SC \ref{infi} we find an infinitesimal presentation for the group $P_n^2$.
In \SC \ref{affine2} we describe the
consequences of considering motions of points in $\P^2$ as opposed to $\A^2$.

The remaining  sections of this paper are devoted to the proof of
Theorem \ref{main theorem1}. In \SC \ref{conj} we see how to
conjugate the generators of $\pi_1(GF_n^2)$ by the generators of $P_{n-1}^2$.
In \SC \ref{recip} we describe a move within the group  $P_n^2$, called
the {\it reciprocity law}, which we use to lift relations
from $P_{n-1}^2$ to $P_n^2$.
This manoeuver is justified in \SC \ref{calc}.
Finally, in \SC \ref{lift} we explain how to lift relations from
$P_{n-1}^2$ into $P_n^2$.

\noindent{ \bf Acknowledgement. }This paper is a shortened version of my
doctoral thesis
which was written at Duke University under the supervision of Professor Richard
Hain.
I would like to thank Professor Hain for his ideas, help and encouragement in
the preparation
of this paper. I want to especially thank him for helping me
understand and carry out the cohomological and infinitesimal
computations which appear in Sections \ref{main} and \ref{infi}.

\s {Affine Versus Projective} \label{affine}

Our definition of $X_n^m$ involved looking at points in $\A^m$. By thinking of
$\A^m$ as being the affine part of $\P^m$, we may extend our definitions to
motions of points in $\P^m$.

Let us first consider the classical braid groups. We denote the classical
pure braid group of $n$ strings on  $\P^1$ by $Q_n$. Since $\P^1$ is
homeomorphic to the two sphere we also see that $Q_n$ is
the classical pure braid group on the sphere. By considering $\C$ to be the
affine part
of $\P^1$ we get a surjective map $P_n \to Q_n$.
Note that the following relations, which do not hold in $P_n$, hold
in $Q_n$:
\begin{eqnarray} \label{qprod}
a_{12}\,a_{13} \dots a_{1n} &  = & 1; \nonumber \\
a_{1k}\,a_{2k}\,\dots \,a_{k-1,k}a_{k,k+1}\,\dots a_{kn} & = & 1  \text{\,\,
for \,\, } 2 \le k \le n.
\end{eqnarray}
(for example see Figure \ref{undo}). In fact it is not hard to show
that these are all
of the extra relations required in the presentation of $P_n$ to get
a presentation of $Q_n$. It is also interesting to note
that the introduction of these relations into the presentation of $P_n$
introduces torsion (e.g. the center of
$Q_n$ contains an element of order $2$---see Corollary \ref{center}).

\begin{figure} \label{undo}
\vskip 1in
\caption{The product $a_{12}a_{13}a_{14}$ is trivial in $Q_4$}
\end{figure}

We now generalise these notions to our situation. Let $Y_n^m$ be the space of
ordered $n$-tuples
in $\P^m$, with $n \ge m+1$
so that each $m+1$ of the points of
each n-tuple span the whole of  $\P^m$.
As with $X_n^m$, the symmetric group on $n$ letters
acts fixed point freely on $Y_n^m$ by permutating the components of each point.
Define $C_n^m = \pi_1(Y_n^m/\Sigma_n)$ and $Q_n^m = \pi_1(Y_n^m)$.
Call these groups {\it the group of n stringed vector braids on $\P^m$}
and {\it the pure group of n stringed vector braids on $\P^m$} respectively.
Note that we have natural maps $P_n^m \to Q_n^m$ and $B_n^m \to C_n^m$.

\begin{lemma} \label{onto}
The natural map $P_n^m \to Q_n^m$ is surjective.
\end{lemma}
\begin{pf}
This follows as  $X_n^m$ is a Zariski open subset
of the smooth variety $Y_n^m$.
\end{pf}
\begin{corollary}
The natural map $B_n^m \to C_n^m$ is surjective. \qed
\end{corollary}

We now define an action of the affine and projective linear groups on $X_n^m$
and $Y_n^m$.

The affine group\footnote{The affine group  $AGL_{m+1}(\C)$ is defined to be
the
stabilizer of the line at infinity in $PGL_{m+1}(\C)$. It is not difficult to
show that
$AGL_{m+1}(\C)$ is the semidirect
product of $GL_{m}(\C)$ by $\C^{m}$.} $AGL_{m+1}(\C)$  acts on the space
$X_n^m$ via the diagonal
action. If $n \ge  m+1$, the the isotropy group of a point is trivial. If
$n=m+1$ then $AGL_{m+1}(\C)$ acts transitively. It follows that $X_{m+1}^m$ is
diffeomorphic to $AGL_{m+1}(\C)$. Let  $\overline{X_n^m}$ denote the quotient
space
$X_n^m / AGL_{m+1}(\C)$. Then $AGL_{m+1}(\C) \to X_n^m \to \overline{X_n^m}$ is
a principal $AGL_{m+1}(\C)$
bundle. Moreover, it has a section (cf. \cite[page 421]{hain-macp}).
Hence we have the following result.

\begin{lemma} \label{split1}
The space $X_n^m$ is diffeomorphic to $\overline{X_n^m} \times AGL_{m+1}(\C)$.
\qed
\end{lemma}

\begin{corollary} \label{center1}
The group $P_n^m$ has a central element of infinite order.
\end{corollary}
\begin{pf}
We have $\pi_1(X_n^m) \cong \pi_1(\overline{X_n^m}) \times
\pi_1(AGL_{m+1}(\C))$. But
$\pi_1(AGL_{m+1}(\C))$ is isomorphic to $\Z$.
\end{pf}

The group $PGL_{m+1}(\C)$ acts on the space $Y_n^m$ via the diagonal
action. If $n \ge  m+2$ the the isotropy group of a point is trivial. If
$n=m+2$ then $PGL_{m+1}(\C)$ acts transitively. It follows that $Y_{m+2}^m$ is
diffeomorphic to $PGL_{m+1}(\C)$. Let  $\overline{Y_n^m}$ denote the quotient
space
$Y_n^m / PGL_{m+1}(\C)$. Then $PGL_{m+1}(\C) \to Y_n^m \to \overline{Y_n^m}$ is
a principle $PGL_{m+1}(\C)$
bundle. It has a section (cf. \cite[page 421]{hain-macp}). Hence we have the
following result.

\begin{lemma} \label{split}
The space $Y_n^m$ is diffeomorphic to $\overline{Y_n^m} \times PGL_{m+1}(\C)$.
\qed
\end{lemma}

\begin{corollary} \label{center}
The group $Q_n^m$ has a central element of order $m+1$.
\end{corollary}
\begin{pf}
We have $\pi_1(Y_n^m) \cong \pi_1(\overline{Y_n^m}) \times
\pi_1(PGL_{m+1}(\C))$. But
$\pi_1(PGL_{m+1}(\C))$ is isomorphic to $\Z/(m+1)\Z$.
\end{pf}

We shall see later that the $AGL_{m+1}(\C)$ action on $X_n^m$ and the
$PGL_{m+1}(\C)$
action on $Y_n^m$ are useful in understanding some of the properties of the
groups $P_n^m$ and  $Q_n^m$.

\s {Getting Started} \label{action}

In this section we study the groups $P_n^m$, $Q_n^m$, $B_n^m$ and $C_n^m$ when
$n \le m+2$.

\begin{proposition} \label{triv}
The groups $P_m^m$ and $Q_{m+1}^m$  are trivial for all $m \in \N$.
\end{proposition}

\begin{pf}
We begin by extending the definition of the space $X_n^m$ to the case where $n
\le m$.
Let $X_n^m$, $1 \le n \le m-1$, be the space of $n$-tuples of points
in $\A^m$ such that the $m$-tuple spans an affine subspace of maximal
dimension.
The map $X_n^m \to X_{n-1}^m$ obtained by forgetting the
last point in each $n$-tuple is a fibration. The fiber of this
map is equal to $\A^m$ less an affine subspace of complex codimension
$m-(n-1)$.
Hence, the fiber of each each of these maps has trivial fundamental
group when $n<m$. Since $X_1^m$ is diffeomorphic to $\A^m$ for each $m$, we can
use
the exact sequence of a fibration to inductively show that, for fixed $m$,
$X_m^m$ has trivial homotopy groups when $n<m$

A similar argument applies to the space $Y_n^m$, giving us the result for
$Q_{m+1}^m$.
\end{pf}

\begin{corollary}
The natural homomorphisms
$$
B_{m}^m \to \Sigma_{m} \text{ \ and \ } C_{m+1}^m \to \Sigma_{m+1}.
$$
are isomorphisms.
\qed
\end{corollary}

The fact that the group $Q_{m+1}^m$ is trivial and
$P_{m+1}^m$ is not is because the space $\A^m$ less a hyperplane is homotopic
to $S^1$,
whereas the space $\P^m$ less a hyperplane is contractible.

\begin{proposition} \label{Z}
The group $P_{m+1}^m$ is isomorphic to $\Z$ for all $m \in \N$.
\end{proposition}
\begin{pf}
The map $X_{m+1}^m  \to X_m^m$ is a fibration, with fiber equal to
$\C^2$ less a line. This is a $K(\Z,1)$. Since $\pi_i(X_m^m)$, is
trivial for $i \ge 1$ (Proposition \ref{triv}), we obtain the result from the
long exact sequence of a fibration.
\end{pf}

We immediately get a similar result for $Q_{m+2}^m$ as a consequence of Lemma
\ref{split}.

\begin{proposition} \label{z/n}
The group  $P_{m+2}^m$ is isomorphic to  $\Z/(m+1)\Z$ for all $m \in \N$. \qed
\end{proposition}

We now use the $PGL_{m+1}(\C)$ action on the space $Y_n^m$ to find a
presentation of the group $C_{m+2}^m$.

\begin{proposition} \label{pressy}
The group $C_{m+2}^m$ admits a presentation with generators
$$
\sigma_1, \dots ,\sigma_{m+1}, \tau,
$$
and defining relations
\begin{eqnarray}
\sigma_i\sigma_j & =  & \sigma_j\sigma_i, \,\,\,\,  |i-j| > 2; \nonumber \\
\sigma_i\sigma_{i+1}\sigma_i & = & \sigma_{i+1}\sigma_i\sigma_{i+1}; \nonumber
\\
\sigma_i \tau & = & \tau \sigma_i; \nonumber \\
\sigma_{i}^2 &  =  & \tau; \nonumber \\
\tau^{m+1} & = & 1. \nonumber
\end{eqnarray}
\end{proposition}

\begin{pf}
In \SC \ref{affine} we saw that the spaces  $Y_{m+2}^m$  and $PGL_{m+1}(\C)$
are diffeomorphic.
Let $e_i$, $1 \le i \le m+1$
be the standard basis  of $\C^{m+1}$. Let $g \in PGL_{m+1}(\C)$.
Then  the map
$$
\theta : PGL_{m+1}(\C) \to Y_{m+2}^m
$$
$$
\theta : g \longmapsto (ge_1,\dots ge_{m+1},ge_1+\dots +ge_{m+1})
$$
is a diffeomorphism.
We know that $\Sigma_{m+2}$ acts on the right of the space $Y^m_{m+2}$ by
permuting coordinates.
We now see that $\Sigma_{m+2}$ also acts on the right of $PGL_{m+1}(\C)$. We do
this by embedding $\Sigma_{m+2}$
into $PGL_{m+1}(\C)$.

Let $s_i$, $1 \le i \le m+1$ denote
the transpositions $(i,i+1)$. Then $\Sigma_{m+2}$ has presentation
\begin{center}
$ \langle s_1,..,s_{m+1}\,|\, s_is_j = s_js_i, |i-j| > 2, \,
s_is_{i+1}s_i =s_{i+1}s_is_{i+1}, 1 \le i \le m+1 \rangle $
\end{center}
Let $P_i \in PGL_{m+1}(\C)$
denote the coset of the permutation matrix corresponding to the transposition
$s_i$ (i.e. the identity matrix with its $i$th and
$(i+1)$st columns swapped).  Map the element $s_i$ to $P_i$ for $1 \le i \le
m$.
Let $P_{m+1} \in PGL_{m+1}(\C)$ be the coset of the matrix
$$
\left[
\begin{array}{ccccc}
 1      &  0      & \cdots & 0      & -1     \\
 0      &  1      & \cdots & 0      & -1     \\
 \vdots &  \vdots & \ddots & \vdots & \vdots \\
 0      &  0      & \cdots & 1      & -1     \\
 0      &  0      & \cdots & 0      & -1
\end{array}
\right]
$$
Map the element $s_{m+1}$ to $P_{m+1}$. Matrix computations show that this map
may be extended to an injective homomorphism from $\Sigma_{m+2}$ to
$PGL_{m+1}(\C)$.

Let $\Sigma_{m+2}$ act on the right of $PGL_{m+2}(\C)$ by group multiplication.
Note that
\begin{eqnarray}
\theta(gP_i) & =  & (gP_i e_1,\dots,gP_i e_{m+1}, gP_i(e_1+\dots +e_{m+1}))
\nonumber           \\
             & =  & (ge_{s_i(1)},\dots,ge_{s_i(m+1)},ge_{s_i(1)}+\dots
+ge_{s_i(m+1)}) \nonumber \\
             & =  & \theta(g)s_i \nonumber
\end{eqnarray}
when $1 \le i \le m$. This is because $P_i e_j$ is equal to the $j$th column of
$P_i$.
Also
$$
\theta(gP_{m+1}) = (e_1,\dots, e_m, -g(e_1 +\dots + e_{m+1}), -ge_{m+1}).
$$
Thus the map $\theta$ is $\Sigma_{m+2}$ equivariant and $PGL_{m+2}(\C)
/\Sigma_{m+2}$ is diffeomorphic
to $Y_{m+2}^m /\Sigma_{m+2}$.

Note that  that $PGL_{m+2}(\C)$ is
isomorphic to $PSL_{m+1}(\C)$. Consequently $SL_n(\C)$ is
the universal cover of  $PGL_n(\C)$.
The natural homomorphism $ SL_{m+1}(\C)  \stackrel{\pi}{\to} PSL_{m+1}(\C)$ is
a $\Z/(m+1)\Z$ covering
whose kernel is generated by the diagonal matrix $\tau$ whose diagonal
entries are all equal to $\exp(2\pi i/(m+1))$.
Let $G = \pi ^{-1}(\Sigma_{m+2})$. Then we have the short exact sequence
$$
1 \to \Z/(m+1)\Z \stackrel{\pi}{\to} G \to \Sigma_{m+2} \to 1.
$$
We use this now to  show that $G$ is given by the same presentation as that
stated in the theorem.
Let $ \omega = \exp(2\pi i/2(m+1))$ and $\sigma_i := \omega P_i$, $1 \le i \le
m+1$.
Then each of these matrices lies in $SL_{m+1}(\C)$.
Use the matrix  $\sigma_i$ as a lift of $P_i$ for $1 \le i \le m+1$ and
the matrix $\tau$ as the generator of the image of $\pi$.
Simple matrix calculations show that the stated relations between the matrices
$\sigma_i$, $1 \le i \le m+1$, and $\tau$ hold in G.

To complete the proof we have to show that $G$ is isomorphic to $C_{m+2}^m$.
First note  that since $\Z/(m+1)\Z$ is central in $G$ we have the isomorphism
$$
SL_{m+1}(\C)/G \cong [\Z/(m+1)\Z \setminus SL_{m+1}(\C)] / \Sigma_{m+2}.
$$
But $[\Z/(m+1)\Z] \setminus SL_{m+1}(\C) $ is isomorphic to $PSL_{m+1}(\C)$,
which is in turn
isomorphic to $Y_{m+2}^m$. Hence
$SL_{m+1}(\C) /G $ is isomorphic to $Y_{m+2}^m /\Sigma_{m+2}$. Since
$SL_{m+1}(\C)$
is the universal cover of $SL_{m+1}(\C) /G $, we conclude that
$$
G \cong \pi_1(Y_{m+2}^m /\Sigma_{m+2}) \cong C_{m+2}^m.
$$
\end{pf}

\s {Forgetting a point}\label{forget}

Let $ p_n^m : X_n^m \to X_{n-1}^m $
denote the map which takes $n$-tuples in $X_n^m$ to $(n-1)$-tuples
in $X_{n-1}^m$ by forgetting the $n$th point. In \cite{F-N}
it is shown that the map $p_n^1$ is a fibration for all $n \ge 2$.
However, in general these maps fail to be fibrations when
$m$ is greater than one. For example, in the case
$m=2$ we have the following result.

\begin{proposition} \label{fibration}
The map $ p_n^2 $ is not a fibration for any  $ n \ge 5$.
\end{proposition}

\begin{pf}
First consider the case when $n=5$. Let $(x_1,\dots ,x_n) \in X_n^2$
and let $L_{ij}$ denote the line
through the points $x_i$ and $x_j$, in the fiber of $p_{n+1}^2$ over
$(x_1,\dots ,x_n)$.
Then the fiber over the point $ (x_1,\dots ,x_n)$ will be equal to $\A^2$ less
the union of
the lines $L_{ij}$. The homotopy type of the fibers will not be constant  since
some
of the fibers will contain parallel lines, which do not intersect in $\A^2$
(see Figure \ref{par} for a non-generic fiber of the map $p_5^2 : X_5^2 \to
X_4^2$).
This problem occurs for all $n \ge 5$.

In the case when $n \ge7$ another type of degeneration
also occurs. We refer to Figure \ref{picnongen}. This is a real picture of the
fibers of $p_n^2$
which shows how one may get non-generic fibers. When lines with disjoint
indices intersect only in double points we are in the generic situation.
However, we see in Figure \ref{picnongen}  that as the line $L_{ij}$
moves upwards it passes through a double point giving us a triple intersection.
The homotopy type of the fiber changes when we obtain triple
points; we no longer have a fibration.
\end{pf}

\begin{figure}
\vskip 1in
\caption{Parallel problems} \label{par}
\end{figure}

Given that the map $p_n^2$ fails to be a fibration when $n \ge 5$,
it might seem hopeless to use the same method that we used
in \SC \ref{intro} to find a presentation of the pure braid group in finding a
presentation for the group $P_n^2$.
However, we are able to partially salvage this situation as follows.

\begin{proposition} \label{p-n-m}
The map $p_n^m$ is topological fibration except over a subset of $X_{n-1}^m$ of
complex codimension one.
\end {proposition}

\begin{pf}
The map $p_n^m$ is a surjective algebraic map, being the
restriction of a projection map  from $(\A^m)^n \to (\A^m)^{n-1}$. The fibers
are complements
of $\left( \begin{array}{c}  \!\!\!n\!\!\! \\  \!\!\!m\!\!\! \end{array}
\right)$ hyperplanes
in $\A^m$. The combinatorics of the
fibers are constant over a Zariski open subset of $X_{n-1}^m$, and
hence the map $p_n^m$ is a topological fibration over this set (a proof of this
fact may be found in
\cite{yau}).
\end{pf}

Using the results contained in \cite{leib} we immediately
obtain.

\begin{corollary} \label{exseq}
Let $GF_n^m$ denote the generic fiber of the map $p_n^m$, $m \ge 1$.
The sequence
$$
\pi_1(GF_n^m) \to \pi_1(X_n^m) \to \pi(X_{n-1}^m) \to 1
$$
\noindent is exact. \qed
\end{corollary}

In the case $m=2$ we have the exact sequence
\begin{equation} \label{seq1}
\pi_1(GF_n^2) \to P_n^2 \to P_{n-1}^2 \to 1,
\end{equation}
\noindent for each $n \ge 4$. This sequence is analogous to the
short exact sequence
$$
1 \to L_{n-1} \to P_n \to P_{n-1} \to 1,
$$
\noindent where $n \ge 3$, involving the classical pure braid groups.
To  date, I have not been able
to prove that the group $\pi_1(GF_n^2)$ injects into $P_n^2$.
If this were the case then Conjecture \ref{hope} would be true.

We close this section by noting that we also have the map $q_n^m : Y_n^m \to
Y_{n-1}^m$,
obtained by forgetting the $n$th point. Both Propositions \ref{fibration} and
\ref{p-n-m}
are also true for the map $q_n^m$. In fact, we can say slightly more in this
case.

\begin{proposition} \label{fibration1}
The map $ q_n^2 : Y_n^2 \to Y_{n-1}^2 $ is a fibration for $n=4,5$ and $6$.
However, the map $ q_n^2 $ is not a fibration for any  $ n \ge 7$.
\end{proposition}

\begin{pf}
First consider the cases when $n$ is equal to $4,5$ and $6$. Let $(x_1,\dots
,x_n) \in X_n^2$
and let $L_{ij}$ denote the line
through the points $x_i$ and $x_j$, in the fiber of $p_{n+1}^2$ over
$(x_1,\dots ,x_n)$.
Then the fiber over the point $ (x_1,\dots ,x_n)$ will be equal to $\P^2$ less
the union of
the lines $L_{ij}$. When $n=5$ and $n=6$, the lines $L_{ij}$ and $L_{kl}$ only
intersect
in a double point when $\{i,j\} \cap \{k,l\} = \emptyset$. Hence, since we are
fibering
over a connected manifold and the combinatorics of all of the fibers are the
same.
The proof now follows from results contained in \cite{yau}.

In the case when $n \ge7$ we refer to Figure \ref{picnongen}.
\end{pf}

\begin{figure}
\vskip 1in
\caption{How degenerate fibers can occur} \label{picnongen}
\end{figure}

\begin{lemma} \label{exact}
Let $F_n^2$ denote the fiber of the map $q_n^2 : Y_n^m \to Y_{n-1}^2$.
For $n=5$ and $n=6$ the sequence
$$
 1 \to \pi_1(F_n^2) \to Q_n^2 \to Q_{n-1}^2 \to 1.
$$
is exact.
\end{lemma}

\begin{pf}
Since $Y_4^2$ is diffeomorphic to $PGL_{m+1}(\C)$ the group
$\pi_2(X_4^2)$ vanishes. Thus, the case $n = 5$ is an immediate consequence of
Proposition \ref{fibration}  and the long exact sequence of fibration.

The case $n=5$
also yields the exact sequence
$$
  \pi_2(F_5^2) \to \pi_2(Y_5^2) \to \pi_2(Y_4^2).
$$
\noindent We now show that $F_5^2$ is a $K(\pi,1)$ space, which implies that
the group
$\pi_2(X_5^2)$ vanishes.
Consider the pencil of lines in $F_5^2$, through the point $b$ (see Figure
\ref{pencil}). This pencil fibers
$F_5^2$. The base is $\P^1$ less $3$ points and the fiber is $\P^1$ less
$4$ points. Hence $F_5^2$ is a  $K(\pi,1)$ space.
\begin{figure}
\vskip 1in
\caption{} \label{pencil}
\end{figure}

By applying Proposition \ref{fibration} and using the long exact
sequence of a fibration once more we obtain the result for the case $n=6$.
\end{pf}

\s {Getting Around} \label{around}

In this section we find generators for the fundamental
group of the complement of a set
of complexified real lines in $\A^2$. Some of the material in this \GS is
drawn from \cite{rand3}.

We begin by stating some conventions that we will use from now on. Fix a real
structure on $\A^n$. We denote the real points of this structure by $\A^n(\R)$.
If $V$ is an affine linear subspace of $\A^n$ then let $V(\R)$ be
equal to $V \cap \A^n(\R)$.

Let $\{ L_i(\R) \}$ be a set of {\it oriented} lines in $\A^2(\R)$.
Denote the union of the $L_i(\R)$ by $\cal{A}(\R)$. Let $(x_1,x_2)$ be
coordinates for $\A^2(\R)$. Then
$(x_1 + iy_1,x_2 + iy_2)$, $y_1,y_2 \in \R$,  are coordinates for $\A^2$.
Once we have chosen coordinates
for $\A^2$ we will abuse notation and write $\C^2$ for $\A^2$ and $\R^2$
for $\A^2(\R)$. Let $\epsilon > 0$ be an arbitrary real number.
It will be convenient to  work in the tubular neighborhood
$N_{\epsilon} = \{ (x_1+iy_1,x_2+iy_2) \in \C^2 \, | \, y_1^2 +y_2^2 \le
\epsilon \} - \cal{A}$ of $\R^2 - \cal{A}(\R)$.
To justify this we require the following proposition.

\begin{proposition}
For all $\epsilon > 0 $, the set  $N_{\epsilon}$ is homotopy equivalent  to
$\C^2 - \cal{A}$.
\end{proposition}
\begin{pf}
We prove this fact using stratified Morse theory \cite{G-M}.
Begin by  stratifying the space $\C^2$ using the arrangement $\cal{A}$ (for
details see \cite[page 245]{G-M}).
Let $f : \C^2 \to \R$ be the function defined by the formula
$f(x_1+iy_1,x_2+iy_2) = y_1^2 + y_2^2$.
This is a Morse function (in the sense of stratified Morse theory) on the
stratified space $\C^2$.
Let $X_{>t}$ denote the set of points $x$  in $\C^2 -\cal{A}$ such that $f(x) >
t$.
Outside the set of points $x \in \C^2$ where $f(x) < \epsilon$ the function $f$
has no critical points.
Hence the set $X_{>t}$ is  homotopy equivalent  to the set $N_{\epsilon}$ for
any $t \ge \epsilon$.
\end{pf}

Let $L(\R)$ be a
generic, oriented line in $\R^2 - \cal{A}(\R)$ and
let $v$ be a vector which orients $L(\R)$.
Its complexification has a canonical orientation and the orientation of the
frame formed by the vectors $v$ and $iv$
agrees with this orientation.
Let $a$ be any point in $L(\R)$. Then we let $\tilde{a} \in L$ denote the point
which is distance $\epsilon$ from $a$ in the direction $iv$.
Let $p \in L(\R) - \cal{A}(\R)$ and $q \in L(\R) \cap \cal{A}(\R)$.
We now define a loop in $N_{\epsilon}$ based at the point $p$ (see Figure
\ref{loop}).
First, move in the direction $iv$ from point $p$ to point $\tilde{p}$. Then
move along the real line in  $L$,
which joins $\tilde{p}$ and
$\tilde{q}$ towards $\tilde{q}$.
On reaching the point $\tilde{q}$, pick a loop $l_q$ with center $q$
and of radius $\epsilon$ within $L$. Now go around this
loop in the positive direction with respect to the  orientation of $L$.
Finally return to $p$ along the same path taken to $\tilde{q}$
from the point $p$. We call this loop {\it the loop in $L$, based at $p$, which
goes around the point $q$.}

\begin{remark} \begin{em}
The homotopy class of the loop which we have just defined depends only upon the
choices of $L(\R)$, its orientation and the points $p$ and $q$.
\end{em} \end{remark}

\begin{figure}
\vskip 1in
\caption{The loop in $L$, based at $p$, which goes around the point $q$}
\label{loop}
\end{figure}

Denote  the set of points in $L(\R) \cap \cal{A}(\R)$ by $\{q_i\}$. Let
$\zeta_i$ be
the loop in $L$, based at $p$ which goes around $q_i$.
Theorems of Lefschetz and Zariski (cf. Chapter 2 of \cite{G-M})  immediately
imply the following
result.

\begin{proposition} \label{agen}
The set of loops $\{ \zeta_i \}$ generates $\pi_1(\C^2 -\cal{A},p)$. \qed
\end{proposition}

\begin{remark} \begin{em}
If we instead considered $\cal{A}(\R)$ as being an arrangement of lines in
$\P^2(\R)$, then
the loops $\zeta_i$  would generate $\pi_1(\P^2 - \cal{A},p)$ (cf. Lemma
\ref{onto}).
\end{em} \end{remark}

We now need a pair of lemmas which will help us manipulate loops in the
complement of a
set of complexified  real lines in $\A^2(\R)$.
To do this we first define a {\it hop}. Let
$H(\R)$ an arbitrary real line in $\A^2(\R)$. Pick a point $p \in
\A^2(\R)-H(\R)$ near to the line $H(\R)$.
The line $H(\R)$ divides $\A^2(\R)$ into two regions.  Let $q \in \A^2(\R)$
be a point in $\A^2(\R)-H(\R)$ close to $p$ that lies in the connected
component of
$\A^2(\R)- H(\R)$ not containing the point $p$. To hop from the
point $p$ to the point $q$ in $\A^2 - H$, choose a line $L(\R)$ joining the
points $p$ and $q$
and follow a loop in $L$ from $p$ to $q$, in the negative direction with
respect to the
canonical orientation of $L$.

The first lemma is local in nature. Let $p$ be an element of $\A^2(\R)$.
Suppose that
$n$ lines $L_j(\R)$, $1 \le  j \le n$,
pass through the point $p$. Label the lines from $1$ through $n$ in
anticlockwise order.
Choose a line $L(\R)$ passing through $p$, which lies between the lines
$L_1(\R)$ and $L_n(\R)$.
Let $C$ be a small sircle in $\A^2(\R)$ centered at $p$.
Let $a,b \in L(\R)$ denote the two points of intersection of the circle $C$
with $L(\R)$.
(see Figure \ref{loc}).
Let $\gamma$ be the path in $\A^2 - \cup L_{j}$, joining the points $a$ and
$b$,
which is obtained by following the circle $C$ in the
anticlockwise direction and  hopping over each line $L_j(\R)$. Let $\gamma'$ be
the path in
$\A^2 - \cup L_j$, joining points $a$ and $b$, which is obtained
by following the circle $C$ in the clockwise direction and hopping over each
line $L_j(\R)$.

\begin{figure}
\vskip 1in
\caption{ } \label{loc}
\end{figure}

\begin{lemma} \label{hops}
The paths $\gamma$ and $\gamma'$ are homotopic in $\A^2 - \cup L_{j}$, relative
to their endpoints $a$ and $b$.
\end{lemma}
\begin{pf}
Let $v$ be the vector in $L(\R)$
from the point $a$ to the point $p$. Let $P_v$ be the plane $\A^2(\R) + iv$.
Then
the intersection $P_v \cap L_{j}$ is empty for $1 \le j \le n$. Hence, the
intersection $P_v \cap [\cup L_j]$ is empty.

If $p$ is an element of $\A^2(\R)$ then let $\tilde{p}$ denote the point $a+iv$
in $P_v$.
Let $C+iv$ denote the circle in $P_v$ which lies above $C$.
Let $\rho$ be the path obtained by going in direction $iv$ from $a$ to
$\tilde{a}$,
going along $C+iv$ in the anticlockwise direction
and finally by going in direction $-iv$ to $b$.
The path $\rho$ is homotopic to $\gamma$ relative to the points $a$ and $b$.
Let $\rho'$ be the path obtained by going in direction $iv$ from $a$ to
$\tilde{a}$,
going along $C+iv$ in the clockwise direction
and finally by going in direction $-iv$ to $b$.
The path $\rho'$ is homotopic to $\gamma'$ relative to the points $a$ and $b$.

To conclude the proof we see that the paths $\rho$ and $\rho'$ are homotopic.
This is because the  paths contained in $C+iv$ which were used to define $\rho$
and $\rho'$ are homotopic
in $P_v$ relative to $\tilde{a}$ and $\tilde{b}$.
\end{pf}

The second lemma is global in nature.
Let $R$ be a bounded region in contained
in $\A^2(\R)$ which is diffeomorphic to a real closed disc. Also assume that
the boundary
of $R$ is smooth.
Let $\cal{A}(\R)$ be an arrangement of real
lines contained in $\A^2(\R)$ such that each line in $\cal{A}(\R)$ is tranverse
to
the boundary of $R$ and the number of components of $R - \cal{A}(\R)$ is
finite.
Also assume that no three lines in $\cal{A}(\R)$ intersect in a point in $R$
and that
the boundary of $R$ contains none of the multiple points of $\cal{A}(\R)$ (see
Figure \ref{ball}).
Let $M(\R)$ denote the set $R - \cal{A}(\R)$ and let $M$ be the set
$$
\{ u+iv : u \in M(\R) , v \in \R^2 , \text{\, and \,} ||v|| < \epsilon \}.
$$
\begin{figure}
\vskip 1in
\caption{} \label{ball}
\end{figure}
We now define some loops in $M$.
Let $\{ L_{\alpha}(\R)$, $\alpha \in A \}$,
denote the set of line  segments in $M(\R)$, obtained by intersecting
$\cal{A}(\R)$ with the set $R$.
Let  $p$ be a base point of $M(\R)$.
For each line segment $L_{\alpha}(\R)$ we define a loop $l_{\alpha}$ in $M$,
based at $p$, as follows.
Pick a point $q$ on the line segment $L_{\alpha}(\R)$ which lies in between
any two  intersection points. Let $l_{\alpha}(\R)$ be a path
in $R$ joining the point $p$ and $q$ which intersects each line segment
$L_{\alpha}(\R)$ transversely only once
and avoids all intersection points.
We now define the loop $l_{\alpha}$.  Follow the path $l_{\alpha}(\R)$, hopping
over any line segment,
until reaching the line $L_{\alpha}(\R)$. Then choose a line $L(\R)$ passing
through $q$, which is
transverse to $L_{\alpha}(\R)$. Now follow a small
loop in $L$, which encircles $L_{\alpha}$,
in the positive direction with respect to the orientation of $L$. Finally,
return to the point $p$
along the same path which was taken outward from $p$.

\begin{lemma} \label{hyps}
Suppose that there exists  a Morse function\footnote{In the sense of stratified
Morse theory.}
defined on the closure of the set $M$ for which each line segment has
a unique critical point, which is not a double point.
Then the group $\pi_1(M,p)$ admits a presentation with generators $l_{\alpha}$,
$\alpha \in A$ and relations
$$
[l_{\alpha},l_{\beta}] =  1,  \quad \text{ if \,} L_{\alpha}(\R) \cap
L_{\beta}(\R) \neq \emptyset .
$$
\end{lemma}
\begin{pf}
Using stratified Morse theory it is not hard to show that we can find a
presentation
for the group $\pi_1(M,p)$ with a generator $l'_{\alpha}$ for each line segment
$L_{\alpha}$.
Moreover the generator $l'_{\alpha}$ can be chosen to be the composition of
a path, which hops over each line segment once,
from $p$ to $L_{\alpha}(\R)$ and a loop which goes around $L_{\alpha}(\R)$, by
following a small loop in a complex line $L$,
in the positive direction with respect to the orientation of $L$. Also, using
Morse theory again, it can be seen that
each pair of the loops $l'_{\alpha}$ and $l'_{\beta}$ commute only if
$L_{\alpha}(\R) \cap L_{\beta}(\R) \neq \emptyset$.

To complete the proof we show that the loops $l'_{\alpha}$
and $l_{\alpha}$ are homotopic in $M$.
The loops $l_{\alpha}$ and $l'_{\alpha}$ were defined by following a path to
the line segment
$L_{\alpha}(\R)$, which hopped  over any line segment once, and
then by following  a small loop which encircled $L_{\alpha}(\R)$.
Note that $L_{\alpha}(\R)$ intersects any other line segment in at most a
double point.
Thus we  can deform the small loop used in the definition of $l'_{\alpha}$ past
any
intersection points in $L_{\alpha}(\R)$, into the small loop which was used to
define the loop $l_{\alpha}$.
In this way we can deform the loop $l'_{\alpha}$ into a new loop $l''_{\alpha}$
which is based at $p$, hops over every line segment on its way to
$L_{\alpha}(\R)$, and
which follows the same small loop around $L_{\alpha}(\R)$ as the loop
$l_{\alpha}$.
Thus we are reduced to showing that we are able  to deform the path used to
define
the loop $l''_{\alpha}$ into the path which is  used to define $l_{\alpha}$.
This can be done by deforming
the path which defines $l''_{\alpha}$, over the  double points in
$\cal{A}(\R)$, into the path
which defines the loop $l_{\alpha}$.
We are able to do this using Lemma \ref{hops}.
\end{pf}

\s {The Generic Fiber}\label{fiber}

In this section we choose a basepoint for the space $X_n^2$. We then find a
natural presentation for the fundamental
group of the generic fiber, $GF_n^2$, of the map  $p_n^2 : X_n^2 \to
X_{n-1}^2$.

To give us some flexibility later, we  begin by
finding a contractible subset $B$ of $X_n^2(\R)$ which we will use to ``fatten
the base point''.
Standard homotopy theory implies that the inclusion $(X_n^2,\ast)
\hookrightarrow (X_n^2,B)$
induces a canonical isomorphism $\pi_1(X_n^2,\ast) \hookrightarrow
\pi_1(X_n^2,B)$ for
all $\ast \in B$. Hence,
elements of  $\pi_1(X_n^2)$ may be represented by  paths
whose endpoints lie within $B$.

Let $x=(x_1, \dots ,x_n)$ be an element of $X_n^2(\R)$ which is mapped to the
point $(x_1, \dots ,x_{n-1})$ by the map $p_n^2$.
Denote the line in the fiber of $p_n^2$ over  the point $x$, which passes
through
$x_i$ and $x_j$, by $L_{ij}$.
Let $\cal{A}_x$ be equal to the union of the
lines $L_{ij}$, $1 \le i <j \le n-1$.
The fiber of the map $p_{n}^2$ over the point
$(x_1,\dots, x_{n-1}) \in X_{n-1}^2$ is then equal to  $ \A^2 - \cal{A}_x$.
Choose coordinates for $\A^2$. Let
$$\psi : \C \to \C^2$$
$$\psi : t \to (t,t^2)$$
be the rational normal curve. Define the set $B$ to be equal to
$$
\{ (\psi(t_1), \dots ,\psi(t_n)) \, | \, t_i \in \R, \, t_1 \le \dots \le t_n
\}.
$$
Since the curve $\psi(\R)$ is convex, the set $B$ is a subset of  $X_n^2(\R)$.
The set $B$ is
clearly contractible.

The fiber $p_n^2$ over a
point in $B$ is not necessarily generic. When choosing a base point in $X_n^2$
we need to  ensure that this is the case and thus
we impose two extra conditions on points in $B$. To do this we first  define
some new lines
in the fiber of $p_n^2$. Consider the curve $\psi(\R)$ as being contained in
$\P^2(\R)$.
Let $L_k^{\infty}(\R)$, $1 \le k \le n$, be equal to the line in $\R^2$ which
passes through the point
$x_k$ and the point at infinity on  $\psi(\R)$. Orient the line
$L_k^{\infty}(\R)$ in the
direction from $x_k$ which points to the point at infinity. We now specify the
two extra  conditions
that we  will impose on points in $B$;

\begin{itemize}
\item Divide the line $L_k^{\infty}(\R)$, $1 \le k \le n-1$, into three
segments as follows.
Let the first segment be the portion of $L_k^{\infty}(\R)$ between $x_k$ and
the point at infinity on $\psi(\R)$.
Let the second segment be the portion of the line $L_k^{\infty}(\R)$
between the point $x_k \in L_k^{\infty}(\R)$ and the point  $L_{12}(\R) \cap
L_k^{\infty}(\R)$. Let the third segment
be the remaining portion of $L_k^{\infty}(\R)$.
A point $x$ of $B$ satisfies the {\it lexcigon condition} if it satisfies the
following three properties;
\begin{enumerate}
\item The lines $L_{ij}(\R)$ with $1 \le i \le k-1$ intersect the
first  segment of the line $L_k^{\infty}(\R)$ in points which lie above the the
lines $L_{i,j-1}(\R)$ for each $k+2 \le j \le n$.
\item The lines $L_{ij}(\R)$, $1 \le i \le j < k$, intersect the
the second  segment of the line $L_k^{\infty}(\R)$ in lexcigongraphical order
with respect to the
orientation of $L_k^{\infty}(\R)$.
\item The lines $L_{ij}(\R)$, $k+1 \le i < j \le n$, intersect the third
segment of $L_k^{\infty}(\R)$
in reverse lexcigongraphical order with respect to the orientation of
$L_k^{\infty}(\R)$.
\end{enumerate}
\item The line $L_k^{\infty}(\R)$, $1 \le k \le n-1$ and the line $L_{kn}(\R)$
divide
$\A^2(\R)$ into four regions. Two of these regions do not
contain the curve $\psi(\R)$. A point $x$ of $B$ satisfies the {\it double
condition} if the regions not containing
the curve $\psi(\R)$ do not contain any double points of the arrangement
$\cal{A}_x$.
\end{itemize}

\begin{remark} \begin{em}
The lexcigon condition ensures that we do not get parallel lines in the fiber
of $p_n^2$ over $x$ as in
Figure \ref{par}.
The double condition ensures that we only get double points in $\cal{A}_x$
(away from the points $x_k$).
Hence we avoid degenerations in the fiber of $p_n^2$ like the one  illustrated
in Figure \ref{picnongen}. \end{em}
\end{remark}

Define the set $S_n$ to be the set of points contained in $B$ which satisfy the
lexcigon
and triple conditions. Note that $p_n^2(S_n)$ is equal to $S_{n-1}$.

\begin{lemma} \label{cont}
The map $p_n^2 : S_n \to S_{n-1}$ is a fibration whose  fiber is homeomorphic
to $\R$.
\end{lemma}

\begin{pf}
We proceed by induction on $n$. When $n=1$ the set $S_1$ is equal to $\R$.
Assume the result up to $n-1$. Let $(x_1,\dots,x_{n-1}) \in S_{n-1}$.
Let $c$ be the final point to the right of $x_{n-1}$ on $\psi(\R)$, for which
$(x_1,\dots,x_{n-1},c)$ fails to satisfy
both the lexcigon and double conditions.
The set of points, $F$,  to the right  of $c$ on $\psi(\R)$ is homeomorphic to
$\R$. Moreover, the
point $(x_1,\dots,x_{n-1},f)$ is in $S_n$ for all $f \in F$.
The result follows.
\end{pf}

\begin{corollary}
The set $S_n$ is contractible for each $n \ge 1$.
\end{corollary}

Let $p_k^2 : X_n^2 \to X_{n-1}^2$ be the projection which forgets the $k$th
point, for $1 \le k \le n-1$.
The set $S_n$ was chosen specifically so that the following result would be
true.

\begin{proposition}
The fiber of the map $p_n^2$ over any point in $S_{n-1}$
is generic. Moreover, the fiber of the map  $p_k^2$, $1 \le k \le n$,
over any point in $S_{n-1}$ is isomorphic to $GF_n^2$
as an  oriented matroid. \qed
\end{proposition}

Let $F_k$ denote the fiber of the map $p_k^2: X_n^2 \to X_{n-1}^2$ over the a
point in $S_{n-1}$. By
\cite[Theorem 5.3]{bj} we immediately obtain the following result.

\begin{corollary}
The groups $\pi_1(GF_n^2)$ and $\pi_1(F_k)$ have the same presentation for $1
\le k \le n-1$. \qed
\end{corollary}

We can now inductively choose a base point for $X_n^2$.  Choose any point in
$S_1$.
Let $b$
be a fixed point in the set $S_n$, with $b$ lying over the
basepoint previously chosen in $S_{n-1}$.
{}From now on define $b$
to be the base point of $X_n^2$.
See Figure \ref{genpic} for a picture of the generic
fiber over the point $b \in X_4^2$.

\begin{figure}
\vskip 1in
\caption{The generic fiber } \label{genpic}

\end{figure}

We shall now find a presentation\footnote{The point $b_n \in GF_n^2$ is the
$n$th component of the base point contained $S_n$.} for the group
$\pi_1(GF_n^2,b_n)$, where $GF_n^2 = \C^2 - \cal{A}_b$
is the generic fiber over the basepoint $b = (b_1, \dots , b_{n-1}) \in
X_{n-1}$.
We  work in the neighborhood $N_{\epsilon}$ defined in \SC \ref{around}.
Note that since $N_{\epsilon}$ is homotopic to $\C^2 - \cal{A}_b$ the two
groups $\pi_1(GF_n^2,b_n)$ and
$\pi_1(N_{\epsilon},b_n)$ are isomorphic.

We begin by finding generators.
Orient all lines $L_{ij}(\R)$, in the direction from $b_i$ to $b_j$ where $i <
j$.
Let $p_{ij} = L_{ij}(\R) \cap L_n^{\infty}(\R)$.
We define  loop $a _{ijn}$, $1 \le i < j \le n-1$,
to be the loop in $L_n^{\infty}$, based
at $b_n$, which goes around $p_{ij}$ (see Figure \ref{atilde}).
As a consequence of Proposition \ref{agen} we immediately obtain the following
result.

\begin{lemma} \label{fgens}
The set
$\{ a_{ijn} \, | \, 1 \le i < j \le n-1\}$, generates the group
$\pi_1(GF_n^2,b_n)$. \newline \qed
\end{lemma}

We now wish to find the defining  relations amongst the $a_{ijn}$  for the
group $\pi_1(GF_n^2,b_n)$.
We will do this using Van Kampen's Theorem.
We begin by dividing  $\R^2 \subset \C^2$ into three open sets.
Let $\epsilon > 0 $ be a small real number.
Let $T(\R)$ be a real tubular neighborhood of the curve $\psi(\R)$,
with diameter equal to $2\epsilon$. Choose $\epsilon$ small
enough so that $T(\R)$ does not contain any double intersections of
$\cal{A}_b$.
Let $D_{\epsilon}(\R) \subset T(\R)$ be
the disc of radius $\epsilon$ centered at $b_n$.
The curve $\psi(\R)$ divides $\R^2$
into two regions. Call the open region containing
the tangent lines to $\psi(\R)$ the
{\it exterior} region of $\psi(\R)$. The complementary
open region will be called the {\it interior} region of $\psi(\R)$.
Let $E(\R)$ denote the union of the exterior region of $\psi(\R)$ and
the disc $D_{\epsilon}(\R)$. Let $I(\R)$ denote the union of the interior
region of
$\psi(\R)$ and the disc $D_{\epsilon}(\R)$. Note that
$\R^2 = T(\R) \cup E(\R) \cup I(\R)$.
Let $E,I$ and $T$  denote the complexification of each of these sets
respectively, intersected with $N_{\epsilon}$.

Our aim is to find a presentation for
$\pi_1(GF_n^2,b_n)$ by applying Van Kampen's Theorem to the sets $I-\cal{A}_b$,
$E-\cal{A}_b$, and $T-\cal{A}_b$.
To do this, we begin by finding a
presentation of $\pi_1(I-\cal{A}_b,b_n)$.
We first  need to define a
new loop $\tilde{a}_{ijn}$, $1 \le i < j \le n-1$, in $GF_n^2$.
Define the loop $\tilde{a}_{ijn}$
as follows. Pick a point $p_{ij} \in I(\R)$ on the line $L_{ij}(\R)$ which lies
within the disc of radius $\epsilon$ about the point $b_j$. Let the line
$\tilde{L}(\R)$ denote the
real line joining $b_n$ and $p_{ij}$.
We define the loop $\tilde{a}_{ijn}$ to
be the loop in $\tilde{L}$, based at $b_n$,
which goes around $p_{ij}$ (see Figure \ref{atilde}).

\begin{figure}
\vskip 1in
\caption {The loops $a_{ijn}$, $\tilde{a}_{ijn}$ and $a'_{ijn}$} \label{atilde}
\end{figure}

\begin{lemma} \label{pre2}
The group $\pi_1(I- \cal{A}_b,b_n)$ admits a presentation with generators
$$
 \tilde{a}_{ijn}, \quad 1 \le i <j \le n-1,
$$
and defining relations
\begin{equation} \label{e2}
[\tilde{a}_{ijn},\tilde{a}_{rsn}] = 1, \quad 1 \le i<r<j<s \le n-1.
\end{equation}
\end{lemma}

\begin{pf}
Note that the set $I(\R)$ is convex. Hence, the lines $L_{ij}(\R)$ intersect
$I(\R)$ in only one segment.
It is not hard to see that one can find a Morse function defined\footnote{ For
example, in order to define
a Morse function, take a family
of real expanding sets, which eventually exhaust $I$, are based at $p$, and
which grow in following way.
The family first intersects each line segment  $L_{ij}(\R) \cap I(\R)$
tangentially.
It also envelops each intersection point one at a time.}  on the closure of the
set $I - \cal{A}_b$.
To complete the proof apply Lemma \ref{hyps}.
\end{pf}

We now find a presentation for the group $\pi_1(E-\cal{A}_b,b_n)$.
We begin by defining a new loop $a'_{ijn}$, $1 \le i < j \le n-1$, contained
in $GF_n^2$. Let $\psi'(\R)$ be a curve joining $b_n$ and $b_{i-1}$ which is
obtained
by deforming the curve $\psi(\R)$ as follows. Fix the points $b_n$ and
$b_{i-1}$ and push the portion of the curve $\psi(\R)$
lying between these two points away from the curve $\psi(\R)$ into the region
$E(\R)$.
Let  $p_{rs}$ be equal to $ L_{rs}(\R) \cap \psi'(\R)$, $1 \le r < s \le n-1$.
We now define the loop $a'_{ijn}$.
Start at the point $b_n$. On reaching a point $p_{rs}$ hop over the line
$L_{rs}(\R)$.
Run all the way down $\psi'(\R)$, hopping over each point $p_{rs}$,
until reaching the line $L_{ij}(\R)$. Then choose a line $L(\R)$ passing
through $p_{ij}$ which is tranverse to
$L_{ij}(\R)$. Let $l$ be a small loop in $L$, which is oriented in the positive
direction with respect to
the orientation of $L$, and which goes around $L_{ij}$.
Run around the  loop $l$ in the positive direction. Finally, return to $b_n$
along the same path
taken on the outward journey.

\begin{lemma} \label{pre1}
The group $\pi_1(E-\cal{A}_b,b_n)$ admits a presentation with generators
$$
a_{ijn} \, \text{ and } \,  a'_{ijn} \quad 1 \le i < j \le n-1,
$$
and defining relations
\begin{equation} \label{e1}
[a_{ijn},a_{rsn}] = 1, \quad  1 \le r<i<j<s \le n-1;
\end{equation}
\begin{equation} \label{ee1}
[a_{ijn},a'_{rsn}] = 1, \quad 1 \le i<j<r<s \le n-1.
\end{equation}
\end{lemma}

\begin{pf}
Note that the set $E(\R)$ is not convex. In fact, the intersection of
$L_{ij}(\R)$ with
$E(\R)$ consists of precisely two line segments (thus for each line
$L_{ij}(\R)$ we have to add the two generators $a_{ijn}$ and $a'_{ijn}$ to the
presentation of $\pi_1(E-\cal{A}_b,b_n)$).
It is not hard to see that one can  define a Morse function on closure of
the set $E - \cal{A}_b$. Now apply Lemma \ref{hyps} to complete the proof.
\end{pf}

Now we find a presentation for the group $\pi_1(T - \cal{A}_b,b_n)$.
Let $[g_1,\dots,g_n]$ denote the relations
$$
g_1\dots g_n = g_2 \dots g_ng_1 = \dots = g_ng_1 \dots g_{n-1}.
$$

\begin{lemma}
The group $\pi_1(T - \cal{A}_b,b_n)$ admits a presentation with generators
$$
 a_{ijk},\quad \tilde{a}_{ijn}, \quad 1 \le i < j \le n-1,
$$
and defining relations
\begin{equation} \label{e3}
[a_{1jn},\dots ,a_{j-1,jn},\tilde{a}_{j,j+1,n},\dots ,\tilde{a}_{j,n-1,n}] = 1,
\quad  1 \le j \le n-1.
\end{equation}
\end{lemma}

\begin{pf}
The generators $a_{ijn}$ and $\tilde{a}_{ijn}$ can be homotoped into the set $T
- \cal{A}_b$.
To see this deform the paths that were used to define the loops $a_{ijn}$ and
$\tilde{a}_{ijn}$ into the set $T - \cal{A}_b$. This can be
done by deforming these paths over the double points in $\cal{A}_b(\R)$, using
Lemma \ref{hops}.
Then deform the small loops in the definition of the $a_{ijn}$ and $a'_{ijn}$,
past the intersection points in lines $L_{ij}(\R)$,
until they lie in the set $T - \cal{A}_b$.

By making $\epsilon$ small enough we see that
$$
\pi_1( T - \cal{A}_b,b_n) \cong  \coprod_{j} \pi_1(B_{\epsilon}(b_j) -
\bigcup_{i} L_{ij}),
$$
where $B_{\epsilon}(b_i)$ is a complex ball or radius $\epsilon$ with center
$b_j$.
We now refer to Randell's paper \cite{rand}. By considering the ends of the
loops $a_{ijk}$ and $\tilde{a}_{ijk}$
locally in each ball $B_{\epsilon}(b_j)$ we see that  the relations arise from
a
calculation made within the complement of the Hopf link,
$\delta(B_{\epsilon}(b_j)) \cap \bigcup L_{ij}$.
\end{pf}

We now want to write the loops $\tilde{a}_{ijn}$  and $a'_{ijn}$ in terms of
the generators $a_{ijk}$.
We refer to Randell's paper \cite{rand}.
Using the result contained in this paper concerning the Hopf link of a point,
we have the formulas
\begin{equation} \label{j1}
\tilde{a}_{ijn} = A^{-1} a_{ijn} A,
\end{equation}
where,
\begin{equation} \label{j2}
A = a_{i-1,jn} \dots a_{1jn},
\end{equation}
and
\begin{equation} \label{j3}
a'_{ijn} = B^{-1} \, \tilde{a}_{ijn} B,
\end{equation}
where,
\begin{equation} \label{j4}
B =  \tilde{a}_{i,j-1,n} \dots \tilde{a}_{i,i+1,n} a_{i,i-1,n} \dots a_{1,i,n}.
\end{equation}

We are now able to state the main theorem of this section.

\begin{theorem}
The group $\pi_1(GF_n^2,b_n)$ admits a presentation with generators
$$
 a_{ijn}, \quad 1 \le i < j \le n-1,
$$
and defining relations (\ref{e2}) -- (\ref{e3}).
\end{theorem}

\begin{pf}
As we have
seen, $N_{\epsilon} \simeq \C^2 -\cal{A}_b$  can be divided into three sets, $E
- \cal{A}_b, I - \cal{A}_b$
and $T- \cal{A}_b$. We know a presentation for the
fundamental group of each of these sets. Hence we need only understand how they
fit together.
The fundamental group of the intersection $E \cap T$ is a
free group with generators $a_{ijn}$ and $a'_{ijn}$.
The fundamental group of the intersection $I \cap T$ is a
free group with generators $\tilde{a}_{ijn}$. Also,
$E \cap I = D_{\epsilon}$ which is contractible.
Now apply  Van Kampen's Theorem.
\end{pf}

In \SC \ref{affine2} we will need to know a presentation for the group
$\pi_1 (\P^2 - \cal{A}_b,b_n)$, which arises as the fundamental
group of the generic fiber of the projection $q_n^2 :Y_n^2 \to Y_{n-1}^2$.

\begin{lemma} \label{extra}
The group $\pi_1(\P^2 - \cal{A}_b,b_n)$ admits a presentation with the
same generators and relations as $\pi(GF_n^2,b_n)$, and with the additional
relation
\begin{equation} \label{e4}
a_{12n}\,a_{13n}\dots a_{1,n-1,n}\,a_{23n}\dots a_{2,n-1,n}\,a_{34n}\dots
a_{n-2,n-1,n} = 1.
\end{equation}
\end{lemma}
\begin{pf}
Use Van Kampen's theorem to  ``glue'' the line at infinity into $GF_n^2$. Note
that this line is homeomorphic
to $\P^1$; hence the relation (\ref{e4}) .
\end{pf}

\s{Main Theorems}\label{main}

We begin this section by finding generators for the group $P_n^2$.
Let $F_k$ be equal to the fiber\footnote{Note that when $k$ is equal to $n$
then
$F_k$ is equal to $GF_n^2$.} over the base point $b \in S_{n-1}$, of the
projection $X_n^2 \to X_{n-1}^2$, obtained by forgetting the $k$th point, for
$1 \le k \le n$.
Let $L(\R)$ be the tangent line
to $\psi(\R)$ passing through the point $b_k$.
Define loop $a_{ijk}$, $1 \le i < j < k \le n$, to be the loop in $L$, based at
$b_k$,
which goes around $L(\R) \cap L_{ij}(\R)$
(see Figure \ref{picgens}). Define loops $a_{ijk}'$ and $\tilde{a_{ijk}}$ in
$P_n^2$ using formulas
(\ref{j1})--(\ref{j4}) with $n=k$.

\begin{figure}
\vskip 1in
\caption{The generator $a_{ijk}$ of $P_n^2$ } \label{picgens}
\end{figure}

\begin{lemma} \label{gens}
The group $P_n^2$, $n \ge 3$, is generated by the set
$$ \{ a_{ijk} \, | \, 1 \le i < j < k \le n \}.$$
\end{lemma}
\begin{pf}
We proceed by induction.
The group  $P_3^2$ isomorphic  to $\Z$ by Proposition \ref{z/n}, thus
when $n=3$ the result is clear.
Assume the result up to $n-1$. Now use sequence (\ref{seq1}).
The generators $a_{ijk}$, where $1 \le i <j <k \le n-1$, generate $P_{n-1}^2$
by induction,
and clearly lift from $P_{n-1}^2$ to $P_n^2$. By adding the
the generators $a_{ijn}$ we obtain the result.
\end{pf}

We now wish to discover relations amongst the $a_{ijk}$ in order
to find a presentation for the group $P_n^2$. We denote the lexcigongraphical
ordering
on the set of two element subsets of $\{1,\dots,n\}$, by $\prec$.

\begin{theorem} \label{main theorem1}
The following relations hold in $P_n^2$, for $3 \le k \le n$.
\begin{equation} \label{comm1}
[a_{ijk},a_{rsk}] = 1, \quad   1 \le r<i<j<s \le k;
\end{equation}
\begin{equation} \label{comm3}
[a_{ijk},a'_{rsk}] = 1, \quad 1 \le i<j<r<s \le k;
\end{equation}
\begin{equation} \label{comm5}
[\tilde{a}_{ijk},\tilde{a}_{rsk}] = 1, \quad 1 \le i<r<j<s \le k;
\end{equation}
\begin{equation} \label{hopf}
[a_{1jk},\dots ,a_{j-1,jk},\tilde{a}_{j,j+1,k},\dots
,\tilde{a}_{j,k-1,k},a_{jk,k+1},\dots ,a_{jk,n-1}] = 1,\text{\  } 1 \le j \le
k-1;
\end{equation}
\begin{equation} \label{conjs}
a_{ijk}^{-1}a_{rst}a_{ijk} =   \hfill
\end{equation}
$$
\left\{ \begin{array}{ll}
a_{rst}  &rs \prec ij \text{ or } jk \prec st; \\
a_{ijt}\,a_{ikt}\,a_{rst}\,a_{ikt}^{-1}\,a_{ijt}^{-1}   &  jk = rs; \\
a_{ijt}\,a_{ikt}\,a_{jkt}\,a_{ikt}^{-1}\,a_{ijt}^{-1}\,a_{jkt}^{-1}\,a_{rst}\,
a_{jkt}\,a_{ijt}\,a_{ikt}\,a_{jkt}^{-1}\,a_{ikt}^{-1}\,a_{ijt}^{-1}
& ik \prec rs \prec jk;\\
a_{ijt}\,a_{ikt}\,a_{jkt}\,a_{rst}\,a_{jkt}^{-1}\,a_{ikt}^{-1}\,a_{ijt}^{-1} &
ik = rs \text{ or } ij = rs;; \\
a_{ijt}\,a_{ikt}\,a_{jkt}\,a_{ijt}^{-1}\,a_{jkt}^{-1}\,a_{ikt}^{-1}\,a_{rst}\,
a_{ikt}\,a_{jkt}\,a_{ijt}\,a_{jkt}^{-1}\,a_{ikt}^{-1}\,a_{ijt}^{-1}
& ij \prec rs \prec ik; \\
& \mbox{where } k < t \le n.
\end{array} \right.
$$
\end{theorem}

\begin{pf}
This proof proceeds inductively using sequence
(\ref{seq1}). When $n=3$ the relations degenerate, which is correct since
$P_3^2$ is isomorphic to $\Z$.

At the $n$th stage we
split the proof up into three main parts and deal
with each separately in the following sections.
\begin{enumerate}
\item When $k=n$ relations (\ref{comm1}) -- (\ref{hopf})
are simply those coming from the
fiber group. These were established in \SC~\ref{fiber}.
\item When $t=n$ relations (\ref{conjs}) come from conjugation
of generators in the fiber group by the generators
of $P_n^2$. These will be established in \SC \ref{conj}.
\item Relations not dealt with in (1) or (2) come from lifting relations from
$P_{n-1}^2$.
These will be established in \SC~\ref{lift}.
\end{enumerate}
\end{pf}

If sequence  (\ref{seq1}) were short exact, then Theorem \ref{main theorem1}
would give
us precisely the relations required for a presentation of $P_n^2$. This
motivates the following definition.

\begin{definition} \label{pln}
Let $PL_n$ be the group whose presentation has generators $a_{ijk}$, $1 \le i
<j <k \le n$,
and relations (\ref{comm1}) -- (\ref{conjs}) of Theorem \ref{main theorem1}.
\end{definition}

Let $\varphi_n : PL_n \to P_n^2$ be the tautological homomorphism
which takes generators to generators. Then,
$\varphi_n$ is clearly well defined and onto.

We now want to see how close the group $PL_n$ is to being isomorphic to
$P_n^2$. We will
do this by studying the relationship between the integral homology of each of
these groups.
Specifically, we will show that $\varphi_n$ induces an isomorphism on
the first two integral homology groups.
First, we will need some preliminary results concerning the homology
of the space $X_n^2$.

\begin{lemma} \label{restrict}
The restriction map
$$
H^k(X_n^2,\Z) \to H^k(GF_n^2,\Z)
$$
is surjective.
\end{lemma}
\begin{pf}
The group $H^1(X_n^2,\Z)$ contains the classes of the  forms
$\frac{1}{2\pi i} \frac{\text{d} \Delta_{ijk} } { \Delta_{ijk} }$, where
$\Delta_{ijk}$ are
the minors defined in the introduction. The
forms $\frac{1}{2\pi i} \frac { \text{d} \Delta_{ijn} } { \Delta_{ijn} }$
restrict to the forms
$\frac{1}{2\pi i} \frac{ \text{d} L_{ijn} }{ L_{ijn} }$ on $GF_n^2$ whose
$k$-fold wedge products
generate $H^k(GF_n^2,\Z)$ \cite{B}.
\end{pf}

\begin{corollary} \label{inj}
The group $H_k(GF_n^2,\Z)$ injects into $H_k(X_n^2,\Z)$ for all $k \ge 0$. \qed
\end{corollary}

\begin{proposition} \label{guts}
The  Leray spectral sequence in the homology for the map $p_n^2 :X_n^2 \to
X_{n-1}^2$ has the following properties
(here all homology groups have $\Z$ coefficients, unless otherwise stated):
\begin{enumerate}
\item For $p+q \le 2$ the  $E_{p,q}^2$ terms are isomorphic to
\begin{center}
\begin{tabular} {|c|c|c|}   \hline
$H_2(GF_n^2)$ &      $ \ast $                            &   $\ast $         \\
\hline
$H_1(GF_n^2)$ & $H_1(GF_n^2) \otimes H_1(X_{n-1}^2)$     &   $\ast $         \\
\hline
$\Z $         & $H_1(X_{n-1}^2)$                         & $H_2(X_{n-1}^2)$  \\
\hline
\end{tabular}
\end{center}
\item Let $p+q \le 2$. Then the $d_2$ differentials whose images lie in
$E_{p,q}^2$ all vanish, and thus $E^2_{p,q} = E^{\infty}_{p,q}$;
\item The terms $E_{p,q}^2$  are torsion free for $p+q \le 2$.
\end{enumerate}
\end{proposition}
\begin{pf}
We begin by proving statement (1). We will work with the  cohomology
spectral sequence first, and justify the statement in homology later.
Let $\pi$ denote the projection map $p_n^2 : X_n^2 \to X_{n-1}^2$.
Let $\cal{U}$ be an open cover of $X_{n-1}^2$.
Let $ \cal{F}_q$ be the sheaf defined by $\cal{F}_q(U) = H^q(\pi^{-1}(U),\Z)$,
where $U \in \cal{U}$. Then the
$E_2$ term of the Leray spectral sequence in cohomology has
$$
E_2^{p,q} = H^p(X_n^2,\cal{F}_q).
$$
and coverges to $H^{p+q}(X_n^2,\Z)$.
If the map $\pi$ were a fibration then $\cal{F}_q$ would be locally constant
for each $q$ and we would obtain
the result,
since the cohomology groups of the fiber of $\pi$ are
torsion free abelian groups of finite rank \cite{B}.
However this is not the case. To obtain the result we show that
$\cal{F}_q$ is locally constant for $q =1,2$.

Let $x$ be any point in $X_{n-1}^2$. We will
show that we can choose an open ball $B_x \subset X_{n-1}^2$, containing
$x$, so that the group $ H^q(\pi^{-1}(B_x),\Z)$ is isomorphic to
$H^q(GF_n^2,\Z)$ for $q = 1,2$.

Denote the complex dimension of the space $X_{n-1}^2$
by $m$. The fiber over the point $x$ in $X_{n-1}^2$ is equal to $\C^2$ less
a union of lines $L_{ij}$, $1 \le i < j \le n-1$.
Let $D_{ijn} \subset (\C^2)^n$ denote the divisor defined  by the minor
$\Delta_{ijn}$ defined in \SC \ref{intro}.
Then we choose an open ball $B_x$ containing $x$ so that $D_{ijn} \cap
\pi^{-1}(B_x)$ is
homeomorphic to $L_{ij} \times B_x$. Denote the set $L_{ij} \times B_x$ by
$F_{ij}$. Then $\pi^{-1}(B_x)$ is
homeomorphic to a complex ball $M$ of dimension $m+2$ minus the union of the
$F_{ij}$.

Since the combinatorics the space $M - \cup F_{ij}$ and the $GF_n^2$
agree in complex codimensions $1$ and $2$ the groups
$ H^q(\pi^{-1}(B_x),\Z)$ and $H^q(GF_n^2,\Z)$ are isomorphic for $q = 1,2$.

We prove (2) and (3) together, using induction. To begin the induction, note
that
$H_0(X_3^2) =\Z$, $H_1(X_3^2) = \Z$, and $H_2(X_3^2) = 0$.
Since $GF_n^2$ is a complement of lines in $\C^2$, we know
that $H_k(GF_n^2)$ (and $H^k(GF_n^2)$) is torsion free for all $k \ge 0$
\cite{B}. We work first with the Leray  spectral sequence in cohomology with
rational coefficients.
Assume inductively that $H^k(X_{n-1}^2)$ is torsion free for $k = 0,1$ and $2$.
{}From Lemma \ref{restrict} we deduce that
all of the differentials $d_2 : E^{0,q}_2 \to E_2^{2,q-1}$ vanish.
By multiplicativity of the spectral sequence in cohomology, this implies that
the
differential $d _2 : E_2^{1,1} \to E_2^{3,0}$ also vanishes.

Now dualize to consider the homology spectral sequence. Note that all of the
required
differentials vanish when the terms of the spectral
sequence have $\Q$ coefficients. However, since the fiber homology is torsion
free
all of the differentials whose image lie in the fiber homology groups vanish
when the
terms have $\Z$ coefficients.
The only other differential which could be non-zero is $d^2 : E^2_{3,0} \to
E^2_{1,1}$.
However, this has to vanish
since $H_1(GF_n^2) \otimes H_1(X_{n-1}^2)$ is torsion free by induction.
\end{pf}

\begin{corollary}
The first and second integral homology groups of $X_n^2$ are torsion free.
\end{corollary}
\begin{pf}
Since the Leray spectral sequence in homology degenerates at  $E_{p,q}^2$, when
$p+q \le 2$,
the graded quotients of the corresponding filtration of $H_1(X_n^2,\Z)$ and
$H_2(X_n^2,\Z)$ are torsion
free, from which the result follows.
\end{pf}

\begin{remark} \begin{em}
The $E_{\infty}$ term of the Leray spectral sequence in cohomology
induces a decreasing filtration $L^{\bullet}$ on $H^{\bullet}(X_n^2,\Z)$
which we shall call the Leray filtration.
\end{em} \end{remark}

We will also need a technical corollary.

\begin{corollary} \label{surj}
The cup product  $\Lambda^2 H^1(X_n^2,\Z) \to H^2(X_n^2,\Z)$ is surjective.
\end{corollary}
\begin{pf}
As a graded ring,
$E_{\infty}^{p,q}$ is isomorphic to $Gr_L^p H^{p+q}(X_n^2)$, where $Gr_L^p$
denotes the
$p$th graded quotient of the filtration $L^{\bullet}$. From the proof of
Proposition \ref{guts}
we deduce that $\Lambda^2Gr_L^{\bullet}H^1(X_n^2,\Z)$ surjects onto
$Gr_L^{\bullet}H^2(X_n^2,\Z)$,
which implies the result by induction on $n$.
\end{pf}

Another immediate consequence of Proposition \ref{guts} is the following
result.

\begin{corollary} \label{h1}
The sequence
$$
0 \to H_1(GF_n^2,\Z) \to  H_1(X_n^2,\Z) \to H_1(X_{n-1}^2,\Z) \to 0
$$
is short exact.
\end{corollary}
\begin{pf}
This sequence arises from the filtration of $H_1(X_n^2,\Z)$ given by the
$E^{\infty}$ term of
the Leray spectral sequence, plus
the fact that $H_1(GF_n^2,\Z)$ injects into $H_1(X_n^2,\Z)$ by Corollary
\ref{inj}.
\end{pf}

This corollary immediately gives us the following lemma.

\begin{lemma} \label{free}
The group  $H_1(X_n^2,\Z)$ is free abelian group of rank $\left(
\begin{array}{c}  \!\!\!n\!\!\! \\ \!\!\!3\!\!\! \end{array} \right)$.
\end{lemma}
\begin{pf}
Proceed by induction. To begin the induction note that the group
$H_1(X_3^2,\Z)$ is isomorphic
to $\Z$, since
$\pi_1(X_3^2)$ equal to $\Z$ (Lemma \ref{Z}).

Since $GF_n^2$ is the complement in $\A^2$ of $\left( \begin{array}{c}
\!\!\!n-1\!\!\! \\ \!\!\!2\!\!\! \end{array} \right)$ lines,
the group $H_1(GF_n^2,\Z)$ is free abelian of rank $\left( \begin{array}{c}
\!\!\!n-1\!\!\! \\  \!\!\!2\!\!\! \end{array} \right)$.
By induction
 the group $H_1(X_{n-1}^2,\Z)$ is free abelian of rank  $\left(
\begin{array}{c}  \!\!\!n-1\!\!\! \\  \!\!\!3\!\!\! \end{array} \right)$.
The result follows from Corollary \ref{h1} and the fact that
$$
\left( \begin{array}{c}  \!\!\!n-1\!\!\! \\  \!\!\!3\!\!\! \end{array} \right)
+  \left( \begin{array}{c}  \!\!\!n-1\!\!\! \\ \!\!\!2\!\!\!\end{array} \right)
= \left( \begin{array}{c}  \!\!\!n\!\!\! \\  \!\!\!3\!\!\! \end{array} \right).
$$
\end{pf}

Since we do not know  if $X_n^2$ is an Eilenberg-MacLane space, it is necessary
to prove
that the first and second integral homology groups of $X_n^2$ are
equal to those of $P_n^2$. To do this we will use some homotopy theory.
Denote the Eilenberg-MacLane space $K(G,1)$ associated to $G$ by $BG$.
The following result may be deduced from obstruction theory and covering space
theory \cite{spanny}.

\begin{lemma} \label{nonsense}
If  $(X,\ast)$ is a connected, pointed topological space with the homotopy type
of a $CW$ complex and fundamental group $G$, then;
\begin{enumerate}
\item there is a natural map $(X,\ast) \to (BG,\ast)$ which is unique up to
homotopy;
\item the homotopy fiber $U$ of the map in (1) is weakly homotopy equivalent to
the universal cover of $X$.
\end{enumerate} \qed
\end{lemma}

\begin{proposition}
When $k =1,2$ the natural map
$$
H_k(X_n^2,\Z) \to H_k(P_n^2,\Z),
$$
is an isomorphism.
\end{proposition}
\begin{pf}
In this proof all homology groups have integer coefficients. We
denote  $\pi_1(X_n^2)$ by $G$.
By Lemma \ref{nonsense} we know that there is a natural map
$$
(X_n^2,\ast) \to (BG,\ast)
$$
whose homotopy fiber, $U_n^2$, is weakly homotopy equivalent the universal
cover of $X_n^2$.
The $E^2$ term Leray-Serre spectral sequence of this fibration is
\begin{center}
\begin{tabular} {|c|c|c|}   \hline
$H_2(U_n^2)$ &      $H_2(U_n^2,H_1(BG))$       &   $H_2(U_n^2,H_2(BG))$  \\
\hline
$0$     & $0$     &   $0 $         \\ \hline
$\Z  $         & $H_1(BG)$                         & $H_2(BG)$  \\ \hline
\end{tabular}
\end{center}
It follows that $H_2(BG)$ injects into $H_2(X_n^2)$. Since
$H_2(X_n^2)$ is torsion free, so is the group $H_2(BG)$.
Since $H_2(BG)$ is isomorphic to $H_2(G)$
we have  the commutative diagram
$$
\begin{CD}
         \Lambda^2H^1(G)            @=       \Lambda^2H^1(X_n^2) \\
         @VVV                                     @VVV               \\
         H^2(G)               @>i^{\ast}>>          H^2(X_n^2)
\end{CD}
$$
By Lemma \ref{surj} the right hand vertical map is surjective. It follows that
the  map $i^{\ast}$ is surjective. Since
$H_2(G)$ is torsion free we can dualize to get the required result.
\end{pf}

We can now prove the main theorems.

\begin{theorem} \label{main theorem2}
The homomorphism $\varphi_n$ induces an isomorphism between the first integral
homology
groups of $PL_n$ and $P_n^2$.
\end{theorem}
\begin{pf}
First consider the group $H_1(PL_n,\Z)$. Since all of the relations in $PL_n$
are commutators $H_1(PL_n,\Z)$ is a free group of rank $\left( \begin{array}{c}
 \!\!\!n\!\!\! \\ \!\!\!3\!\!\! \end{array} \right)$ generated by the homology
classes of the elements $a_{ijk}$, $i \le i < j < k \le n$.
The map $PL_n \to PL_{n-1}$ induces a map from $H_1(PL_n,\Z) \to
H_1(PL_{n-1},\Z)$.
By considering the abelianisation of the relevant  groups we obtain the
short exact sequence of torsion free abelian groups
$$
0 \to K \to H_1(PL_n,\Z) \to H_1(PL_{n-1},\Z) \to 0
$$
where $K$ is defined to be the kernel.
The group $K$ is generated by the homology classes of the elements $a_{ijn}$,
$1 \le i < j  \le n-1$.

The map $\varphi_n$ induces the following commutative diagram.
$$
\begin{CD}
0       @>>>      K         @>>>    H_1(PL_n,\Z)                 @>>>
H_1(PL_{n-1},\Z)     @>>>        0    \\
&        &       @VV{i}V         @VV{{\varphi_n}_\ast}V
@VV{{\varphi_{n-1}}_\ast}V      &         &    \\
0       @>>> H_1(GF_n^2,\Z) @>>>    H_1(P_n^2,\Z)                @>>>
H_1(P_{n-1}^2,\Z)     @>>>       0
\end{CD}
$$
The map $i$ is clearly well defined and the bottom row of the diagram is exact
by Lemma \ref{h1}.
The map $i$ is surjective since $H_1(GF_n^2,\Z)$ is generated by the homology
classes of the loops
$a_{ijn}$, $1 \le i < j \le n-1$.
Since the groups $K$ and $H_1(GF_n^2,\Z)$ are both torsion free abelian groups
the map $i$ is an isomorphism.

We now proceed by induction. Note that $H_1(PL_3,\Z)$ and $H_1(P_3^2,\Z)$ are
both isomorphic to $\Z$.
Assume by induction that the map ${\varphi_{n-1}}_{\ast}$ is an isomorphism.
The map $i$  is an isomorphism and so we complete the proof by applying the
Five Lemma to the
above commutative diagram.
\end{pf}

To prove that $H_2(P_n^2,\Z)$ is isomorphic to $H_2(PL_n,\Z)$ we will use some
rational
homotopy theory. Let $D : H_2(X,\Q) \to \Lambda ^2H_1(X,\Q)$ be the map induced
by the diagonal inclusion $X \hookrightarrow X \times X$. Given a group $G$,
let $\Gamma^nG$ denote
the $n$th term in the lower central series of $G$.
The proof of the following result is elementary and may be found in
\cite{serre}.

\begin{lemma} \label{ser}
If $G$ is any group then the commutator map
$$
\Lambda^2H_1(G,\Z) \stackrel{[\,\,,\,\,]}{\to} \Gamma^2G/\Gamma^3G
$$
defined by
$$
x \wedge y \longmapsto \overline{ xyx^{-1}y^{-1} }
$$
is a surjection. \qed
\end{lemma}

The following result can be proved using results in either \cite[\S 2.1]{chen}
or \cite[\S 8]{S}.

\begin{lemma} \label{guts2}
If $X$ is a topological space and the dimension of $H_1(X,\Q)$ is finite for $k
=1,2$, then the sequence
$$
H_2(X,\Q) \stackrel{D}{\to} \Lambda^2H_1(X,\Q) \stackrel{[\,\,,\,\,]}{\to}
[\Gamma^2\pi_1(X)/\Gamma^3\pi_1(X)] \otimes \Q \to 0
$$
is exact and natural in $X$. \qed
\end{lemma}

\begin{corollary}
The sequence
$$
0 \to H_2(X_n^2,\Q) \to \Lambda^2H_1(X_n^2,\Q) \to
[\Gamma^2\pi_1(X_n^2)/\Gamma^3\pi_1(X_n^2)] \otimes \Q \to 0
$$
is short exact.
\end{corollary}
\begin{pf}
Note that the map $H_2(X,\Q) \to \Lambda^2H_1(X,\Q)$ in Lemma \ref{guts2} is
injective if and
only if $\Lambda^2H^1(X,\Q) \to H^2(X,\Q)$ is surjective. Now apply Lemma
\ref{surj}
\end{pf}

We need to compute the map $D : H_2(PL_n,\Q) \to \Lambda^2H_1(PL_n,\Q)$.
The following result is proved in \cite{serre}.

\begin{lemma} \label{balls2}
If $F$ is a free group then
$$
\Lambda^2H_1(F,\Z) \stackrel{[\,\,,\,\,]}{\to} \Gamma^2F/\Gamma^3F
$$
is an isomorphism of $\Z$-modules. \qed
\end{lemma}

Let $G$ be the group
$$
G = \langle x_1, \dots ,x_n \,\, | \,\, r \rangle.
$$
where $r \in \Gamma^2\langle x_1, \dots , x_n \rangle$.
The relation $r$ gives  an element of $H_2(G,\Z)$. We will call
$D(r)$ the linearization of $r$.
As a consequence of Lemma \ref{balls2} and Lemma \ref{guts2} we have the
following result.

\begin{lemma}
If $r$ is as above then;
\begin{enumerate}
\item the group $H_2(G,\Z)$ is isomorphic to $\Z[r]$;
\item if $r \equiv \prod [x_i,x_j] ^ {m_{ij}} \text{\, mod \,} \Gamma^3G$, then
the map
$D$ is given by the formula
$$
D(r) = \sum m_{ij} (x_i \wedge x_j).
$$
\end{enumerate} \qed
\end{lemma}

\begin{theorem} \label{main theorem3}
The homomorphism $\varphi_n$ induces an isomorphism between the second integral
homology
groups of $PL_n$ and $P_n^2$.
\end{theorem}
\begin{pf}
Let $N$ be the number of relations in $PL_n$.
The rank of $H_2(PL_n, \Z)$ is less than or equal to $N$. When  equality occurs
$H_2(PL_n,\Z)$ is torsion free. So if  we can show that the rank of
$H_2(PL_n,\Z)$ is equal to the dimension of $H_2(PL_n,\Q)$, then $H_2(PL_n,\Z)$
is
torsion free. Thus we first use rational coefficients.

We begin by defining filtrations on $H_1(PL_n)$, $\Lambda^2H_1(PL_n)$ and
$H_2(PL_n)$.
Let $G_0$ be equal to the set of $a_{ijn}$ where $1 \le i < j \le n-1$, and
$G_1$ be equal to $H_1(PL_n)$. Then
the the subspaces $\text{span}\{G_i\}$ filter $H_1(PL_n)$.
This filtration induces the following filtration on $\Lambda^2H_1$.
Let $F_0 = \text{span}\{a_{ijn}\wedge a_{rsn} \}$. Let $F_1  =
\text{span}\{a_{ijk} \wedge a_{rsn}\} \cup F_0$
where $k \neq n$. Finally, let $F_2 = \Lambda^2H_1(PL_n)$.

To define a filtration on $H_2(PL_n)$ we begin by
filtering the relations of $PL_n$. Let
$R_0$ be the set of relations (\ref{comm1})--(\ref{hopf}) with $k=n$. These
relations arise from the fiber group.
Let $R_1$ be the set of relations
(\ref{conjs}) with $t=n$ together with the set relations $R_0$. The set of
relations $R_1 - R_0$ come from conjugating
generators in the fiber group by generators in $P_{n-1}^2$. Finally, let $R_2$
be the set of all relations for $PL_n$.
We now define a  filtration $\tilde{L}$ on $H_2(PL_n)$.
Let $\tilde{L}_i$ be the span of the elements of $H_2(PL_n)$ coming from the
elements of $R_i$.

We now proceed with the proof using induction. First, note that both
$H_2(PL_3,\Z)$ and
$H_2(P_n^2,\Z)$ are trivial.
Assume the result up to $n-1$.
As a consequence of Lemma \ref{guts2} we have the following commutative
diagram.
$$
\begin{CD}
&        &         H_2(PL_n)             @>>>   \Lambda^2H_1(PL_n)    @>>>
[\Gamma^2PL_n / \Gamma^3PL_n]\otimes \Q   @>>>     0     \\
&        &       @VV{{\varphi_n}_\ast}V               @VV{\cong}V
             @VVV                      &       &    \\
0       @>>>      H_2(X_n^2)             @>>>   \Lambda^2H_1(X_n^2)   @>>>
[\Gamma^2P_n^2 / \Gamma^3P_n^2]\otimes\Q  @>>>     0
\end{CD}
$$
First we show that the map $D : H_2(PL_n) \to  \Lambda^2H_1(PL_n)$ is
injective. The map $D$ is
filtration preserving.
Thus the
map $ \gamma_i : \tilde{L}_i/\tilde{L}_{i-1} \to F_i/F_{i-1}$ induced by $D$ is
well defined. To
show that $D$ is injective it is
sufficient to show that $\gamma_i$ is injective for $0 \le i \le 2$.
We begin with $\gamma_0$. The image under $D$ of the fiber relations are
\begin{equation} \label{lin1}
a_{ijn} \wedge a_{rsn}, \text{ if } ij \neq rs;
\end{equation}
\begin{equation} \label{lin11}
a_{ijn} \wedge (a_{1in} + a_{2in} + \dots  + a_{ijn} +  \dots  + a_{i,n-1,n}),
 \text{ for } 1 \le i \le n-1.
\end{equation}
All of these are linearly independent in $F_0$. Hence, $\tilde{L_0}$ injects
into $F_0$.

Now consider the map $ \gamma_1 $.
The linearized conjugation relations are
\begin{equation} \label{lin2}
a_{ijk} \wedge a_{rsn} ,\text{ if } rs \neq ij, \,  ik \text{ or } jk;
\end{equation}
\begin{equation} \label{lin22}
a_{rsn} \wedge (a_{ijk}+a_{ijn}+a_{ikn}+ a_{ikn}) , \text{ if } rs  =  ij, \,
ik \text{ or } jk .
\end{equation}

%\footnote {These are analogous to the infinitesimal braid relations of Kohno
% \cite[pp ]{k}}

\noindent We now quotient out by $F_0$. Relations (\ref{lin2}) -- (\ref{lin22})
modulo $F_0$
become
\begin{equation} \label{balls}
a_{rsn} \wedge a_{ijk} , \text{\,\,  where \,\,} k \neq n.
\end{equation}
These are linearly independent. Hence $\gamma_1$ is injective.
The map $ \gamma_2$
is injective by induction. Hence the map D is injective.

We now show that
the map $\varphi_n$ induces an isomorphism between $H_2(PL_n)$ and
$H_2(X_n^2)$.
Let $L$ be the filtration on $H_{\bullet}(X_n^2)$ induced by the Leray
sequence. Since the
map $D$ is injective for both $H_2(PL_n)$ and $H_2(P_n^2)$ we will
not differentiate  between $H_2$ and its image in $\Lambda^2H_1$; for example
$D(L_0) = L_0$.  We
show that the map $ \beta_i : \tilde{L}_i/\tilde{L}_{i-1} \to L_i/L_{i-1}$
induced by $\varphi_n$
is an isomorphism for $0 \le i \le 2$.

Begin with the map $\beta_0$. Note that $L_0 = H_2(GF_n^2)$. Thus we have  to
analyse the group
$H_2(GF_n^2)$. Recall that $GF_n^2 = \C^2 - \cal{A}_b$. Let $p$ be an
intersection point in $\cal{A}_b$
and $\cal{A}_p$ be the set of lines in $\cal{A}$ passing
through the point $p$. Let $M_p = \C^2 - \cal{A}_p$. Then there exist inclusion
maps $i_p : M_p \hookrightarrow GF_n^2$.
By a result of Brieskorn \cite{B} the maps $i_p$ induce an isomorphism
$$
\bigoplus_{p} H_2(M_p) \cong H_2(GF_n^2).
$$
Note that each  relation in $\pi_1(GF_n^2)$  arises from an intersection point
$p$ of $\cal{A}_b$.
We are thus reduced to the complement of $n$ lines through the origin in
$\C^2$. Denote this space by
$M$. The dimension of $H_2(M)$ is equal to $n-1$ \cite{orlik}. Now consider the
dimension of the space $\cal{L} \subset \Lambda^2H_1(M)$ spanned by the
linearised relations for $\pi_1(M)$. The linearizations take the form
$$
a_i \wedge (a_1+\dots + a_n), \text{ \, for \,} 1 \le i \le n
$$
where $a_i$ generates $H_1(M)$. However, the sum of all of these is equal to
zero.
Thus the dimension of $\cal{L}$ is equal to $n-1$, as required.

Now consider the map $\beta_1$. The quotient $L_1/L_0$ is isomorphic
to $H_1(GF_n^2) \otimes H_1(X_n^2)$. But this is clearly isomorphic to
$\tilde{L}_1/\tilde{L}_0$
via $\beta_1$ by equation (\ref{balls}).
The map $\beta_2$ is an isomorphism by induction.
We conclude that the map ${\varphi_n}_{\ast}$ is also an isomorphism.

We now complete the proof by considering integer coefficients.
First, since the map  $H_2(PL_n) \to  \Lambda^2H_1(PL_n)$
is injective, the dimension of $H_2(PL_n)$ is equal to $N$. Thus $H_2(PL_n,\Z)$
is torsion free and the
map  $D :H_2(PL_n,\Z) \to  \Lambda^2H_1(PL_n,\Z)$ is well defined. Hence we
have the following commutative diagram.
$$
\begin{CD}
       H_2(PL_n,\Z)             @>>>   \Lambda^2H_1(PL_n,\Z)   \\
  @VV{{\varphi_n}_\ast}V                     @VV{\cong}V       \\
       H_2(X_n^2,\Z)            @>>>   \Lambda^2H_1(X_n^2,\Z)
\end{CD}
$$
Let $I_1 = \text{Im}\{D : H_2(PL_n,\Z) \to \Lambda^2H_1(PL_n,\Z) \}$, and $I_2
= \text{Im}\{D : H_2(P_n^2,\Z) \to \Lambda^2H_1(X_n^2,\Z) \}$.
To complete the proof it suffices to show that $I_1 = I_2$. Note that $I_1
\subset I_2$.
Thus it suffices to show that $I_1$ is primitive in $I_2$, i.e.
$$
I_1 = (I_1 \otimes \Q) \cap I_2.
$$
This can be shown easily using the isomorphism $Gr_{\bullet}^L I_1 \cong
(Gr_{\bullet}^L I_1 \otimes \Q) \cap I_2$.
\end{pf}

\s{Infinitesimal Vector Braid Relations} \label{infi}

We begin by recalling the infinitesimal presentation of the classical pure
braid group.
Given a group $G$ we denote its group algebra over $\C$ by $\C [G]$. Let
$\epsilon: \C [G] \to \C$ be
the augmentation homomorphism and $J$ be equal to the kernal of $\epsilon$.
The powers of $J$ define a topology on $\C [G]$ which is
called the $J$-adic topology.
In what follows we let $\C  \langle Y_i \rangle $ denote the free associative,
non-commutative algebra in the
indeterminants $Y_i$ and $\C \langle \langle Y_i \rangle \rangle$
denote the non-commutative formal power
series ring in the indeterminants $Y_i$.
In \cite{kohno} Kohno proves that
the $J$-adic completion of the group ring
$\C [P_n]$ is isomorphic to $\C \langle \langle X_{ij} \rangle \rangle$ modulo
the two-sided ideal
generated by the relations
$$
[X_{ij}, X_{ik} + X_{jk}] \quad i<j<k;
$$
$$
[X_{jk}, X_{ij} + X_{ik}] \quad i<j<k;
$$
$$
[X_{ij},X_{rs}]  \quad i,j,r,s \quad \text{distinct.}
$$

We now find the corresponding infinitesimal presentation  for the group
$P_n^2$.

\begin{proposition} \label{provin}
The completed group ring of $P_n^2$ is isomorphic to
$\C \langle  \langle X_{ijk} \rangle  \rangle $, $1 \le i < j < k \le n$,
modulo
the two-sided ideal generated by the relations
\begin{equation} \label{in1}
[X_{ijk}, X_{jkl} + X_{ikl} + X_{ijl} ]  \quad i<j<k<l;
\end{equation}
\begin{equation} \label{in2}
[X_{jkl}, X_{ijk} + X_{ikl} + X_{ijl} ]  \quad i<j<k<l;
\end{equation}
\begin{equation} \label{in3}
[X_{ikl}, X_{ijk} + X_{ijl} + X_{jkl} ]  \quad i<j<k<l;
\end{equation}
\begin{equation} \label{in4}
[X_{ijk}, X_{12k} + \dots + X_{k,n-1,n}] \quad 1 \le k \le n;
\end{equation}
\begin{equation} \label{in5}
[X_{ijk},X_{rst}] \quad i,j,k,r,s,t \quad \text{distinct.}
\end{equation}
\end{proposition}
\begin{pf}
This is a special case of a result due to K.-T.~Chen \cite{chen2}. We
apply the version of this result which  appears in \cite[pages 28-29]{hain}.
Observe that the tensor algebra $ \bigoplus\limits_{n =0}^{\infty}
H_1(X_n^2,\C)^{\otimes n}$ on $H_1(X_n^2,\C)$
is isomorphic to $\C \langle X_{ijk} \rangle$, where the indeterminant
$X_{ijk}$ denotes the
homology class of the generator $a_{ijk}$ of the group $P_n^2$. This is a
direct consequence of Lemma \ref{free}
and Theorem \ref{main theorem2}.
Let
$$
\delta : H_2(X_n^2,\C) \to H_1(X_n^2,\C)^{\otimes 2} \subset \C \langle X_{ijk}
\rangle
$$
be the dual of the cup product. According to \cite[pages 28-29]{hain}, the
completed group
ring of $P_n^2$ is isomorphic to $\C \langle \langle X_{ijk} \rangle \rangle /
(\text{im}\, \delta)$.
Using the results contained in the proof of Theorem \ref{main theorem3} the
ideal $(\text{im}\, \delta)$
is simply the two-sided ideal generated by the relations
(\ref{in1})--(\ref{in5}).
\end{pf}

\s{Affine Versus Projective Revisited} \label{affine2}

As we saw in \SC \ref{affine}, we can consider motions of points in $\A^m$
as being motions of points in $\P^m$. This has some interesting consequences
for
the groups $P_n^2$ and $Q_n^2$. First, note that we have the natural surjective
map from $P_n^2 \to Q_n^2$
(see Lemma \ref{onto}).
Thus, we immediately see that $Q_n^2$ is generated by $a_{ijk}$, for $1 \le i <
j < k \le n$.
However, since we are now
looking at points in $\P^2$ we get some extra relations amongst these
generators, which are
analogous to relations (\ref{qprod}) of $Q_n$.

\begin{lemma}
For $1  \le k \le n$, the following relations hold in $Q_n^2$
\begin{eqnarray} \label{prod}
a_{123}\,a_{124}\dots a_{12n} \, a_{134} \dots a_{13n} a_{145} \dots
a_{1,n-1,n}                & = & 1; \nonumber  \\
a_{123}\,a_{124}\dots a_{12n} \, a_{234} \dots a_{23n} a_{245} \dots
a_{2,n-1,n}                & = & 1; \nonumber  \\
a_{12k}\,a_{13k}\dots a_{k-2,k-1,k}\,a_{1k,k+1}\dots a_{k-1,k,k+1}\dots
a_{1kn}\dots a_{k-1,kn} & = & 1.
\end{eqnarray}
\end{lemma}
\begin{pf}
Each of these relations arises from a product relation which occurs in the
fiber of the projection, $Y_n^2 \to Y_{n-1}^2$, obtained by forgetting the
$k$th point.
These can then be
written in the required form using the reciprocity law. See \SC \ref{prods} for
more details.
\end{pf}

We define a group $QL_n^2$ by adjoining the extra relations (\ref{prod}) to the
presentation of $PL_n$, and conjecture that this group is
isomorphic to $Q_n^2$.
An argument similar to that in the proof of Theorem \ref{main theorem2}
can be used to show that the first homology groups of $QL_n$ and $Q_n^2$ are
the same, and are free abelian of rank
$\left( \begin{array}{c}  \!\!\!n\!\!\! \\  \!\!\!3\!\!\! \end{array} \right)
-n$.

We can exploit the action of $PGL_3(\C)$ on $Y_n^2$ to better understand the
group $Q_n^2$.
As we have already seen, the action can be used
to show that the group $Q_n^2$ has a cental element of order three (see Lemma
\ref{center}). Denote
this element by $\tau$. In fact, by
analysing the action of $PGL_3(\C)$ on $Y_n^2$ we can find an explicit formula
for this element in
terms of the generators $a_{ijk}$.

\begin{lemma}
The element $\tau$ is given by the formula
$$
\tau = \tau_3 \dots \tau_n
$$
where
$$
\tau_i = a_{12i}a_{13i} \dots a_{i-2,i-1,i}.
$$
\end{lemma}
\begin{pf}
We analyse where the map  $PGL_3(\C) \to Y_n^2$ sends the generator $\rho: S^1
\to PGL_3(\C)$
of $\pi_1(PGL_3(\C))$,
given by the formula
$$
\rho : \theta \longmapsto \left[
\begin{array}{ccccc}
 1       &  0         &  0       \\
 0       &  1         &  0       \\
 0       &  0         &  e^{i\theta}
\end{array}
\right]
$$
where $0 \le \theta \le 2\pi$.

Choose coordinates $(x,y)$, $x,y \in \R$  for the affine part of
$\P^2(\R)$. Let the points $b_1$ and $b_2$ be
equal to $(0,0)$ and $(0,1)$ respectively.
Denote the coordinate of the  point $b_i$ by  $(x_i,y_i)$, for $i \ge 3$. The
line $L_i^{\infty}(\R)$ is equal to
the vertical line passing through $(x_i,y_i)$.
We ``stretch out'' the points $b_i$ on the curve $\psi(\R)$ so that they
satisfy
the following condition\footnote{Note we may do this whilst remaining within
the
basepoint set $B$ which was defined in \SC \ref{fiber}.}.
We require that the line $L_{i,i+1}(\R)$ intersects
the line $L_{i-1}^{\infty}(\R)$ at a point whose $y$ coordinate is less than
$-y_{i-1}$.

The loop $\rho$ will be sent to the loop $\tau : S^1 \to Y_n^2$ given by the
formula
$$
\tau : \theta \to ((0,0), (0,1), (x_3,e^{i\theta}y_3),
\dots,(x_n,e^{i\theta}y_n)).
$$
On ``squeezing'' the points back to their original position, we see that
the loop which each point $b_i$ follows in the loop $\tau$  is homotopic to the
loop
$$
\tau_i = a_{12i}a_{13i} \dots a_{i-2,i-1,i}.
$$

We now show that the loops $\tau_i$ commute with each other. Fix $i < j$.
Note that the loop $\tau_j$ encircles the lines $L_{rs}$ for $1 \le r < s \le
j-1$. As the  point $b_i$
follows the loop $\tau_i$ the lines $L_{rs}$ always remain within the loop
$\tau_j$. Thus the motion
of the point $b_i$ along  $\tau_i$  is independent of the loop $\tau_j$.

Finally note that since the $\tau_i$ commute we have the expression
$$
\tau = \tau_3 \dots \tau_n,
$$
as required.
\end{pf}

\begin{figure}
\vskip 1in
\caption{Let's do the twist!} \label{twist}
\end{figure}

The element $\tau$ of order three in $Q_n^2$ is
analogous to the full-twist $\Delta$ of the classical braid group of the
sphere.
We recall some results about the classical braid groups \cite{birman}.
In the classical case $P_n$ is a group with center $\Z$ generated by the
full-twist
$$
\Delta = \delta_2 \dots \delta_n,
$$
where
$$
\delta_i = a_{1i} a_{2i} \dots a_{i-1,i}.
$$
The elements $\delta_i$ all commute with each other (see Figure \ref{twist}).
Moreover, $\Delta$ is a central element of $Q_n$ which has order $2$.

Using the classical braid groups as a model, it is natural to conjecture that
$\tau$
generates the center of the group $P_n^2$.

\s{Conjugation}\label{conj}

In this section we describe a method for conjugating generators
of $\pi_1(GF_n^2)$ by generators of the group $P_{n-1}^2$.

It will be helpful to first discuss the classical pure braid
case.  Recall  the short exact sequence (\ref{pureseq}).
To find a presentation for $P_n$ from one for $P_{n-1}$, we have to
be able to write $a_{ij}^{-1} a_{rn} a_{ij}$ as a word in the
group $L_{n-1}$, where $1 \le i < j \le n-1$, and $1 \le r \le n-1$.
We can do this by picturing the loops $a_{rn}$ and $a_{ij}$ in
$\C$ ( Figure \ref{purepic} ).
As we ``unwind'' the braid $a_{ij}^{-1} a_{rn} a_{ij}$,
the point $j$ moves around point $i$ and ``pushes'' the loop
$a_{rn}$ with it. At the end of the unwinding process,
we are left with a  loop in $L_{n-1}$, which will be the required conjugate of
$a_{rn}$. We can describe this loop in terms of the fiber generators
by recording where this loop crosses vertical half lines below the
points $\{1,..,n-1\}$. For example the relation
$$
a_{ij}^{-1} \, a_{rs} \, a_{ij} =
a_{is}\,a_{rs}\,a_{is}^{-1}, \quad 1 \le i<r=j<s \le n,
$$
is shown in Figure \ref{pureconjpic}.

\begin{figure}
\vskip 1in
\caption{The generator $a_{ij}$ of $P_n$} \label{purepic}
\end{figure}

\begin{figure}
\vskip 1in
\caption{Conjugation in $P_n$} \label{pureconjpic}
\end{figure}

We now describe a similar process for conjugating in $P_n^2$.
Let $L(\R)$ denote the line $L_n^{\infty}(\R)$ which we used to define the
loop $a_{rsn}$, $1 \le r < s \le n-1$.
Let $p_{rs}$ denote the intersection
of line $L_{rs}(\R)$ with $L(\R)$. We picture the generator
$a_{rsn}$ contained in $L$ in Figure \ref{aijk}.

\begin{figure}
\vskip 1in
\caption{The generator $a_{rsn}$ of $P_n^2$} \label{aijk}
\end{figure}

Note that if we move the point $b_k$ for $1 \le k \le n-1$,
then we will induce a motion of the points $p_{rs}$ within $L$.
Hence, when point $b_k$ follows the loop $a_{ijk}$, we can picture
the induced motions of the points $p_{rs}$ in $L$. We can
use this picture to  determine how
generator $a_{ijk}$ conjugates generator $a_{rsn}$ by recording how
the movements of the $p_{rs}$ within $L$ deform the loop $a_{rsn}$.

At this point we need to make some observations which will simplify the
calculation. First, note that the motion of the point $b_k$ only
induces  a motion of the points $p_{rk}$. Moreover, by choosing the
loop $l$
which the point $b_k$ follows on $a_{ijk}$ whilst going around $L_{ij}(\R)$ to
be small
enough\footnote{See the definition of $a_{ijk}$.}, we can ensure that only the
points $p_{jk}$ and $p_{ik}$ go
around  $p_{ij}$ as the point $b_k$ goes around $l$.
Hence, we need only consider the motion of the points $p_{ik}$ and
$p_{jk}$ within $L$. We picture the motion of these points induced by the
motion of $b_k$
in Figure \ref{motion}.

\begin{figure}
\vskip 1in
\caption{Induced motion} \label{motion}
\end{figure}

In Figure \ref{movie} we picture how to obtain the relation
$$a_{ijk}^{-1}a_{rsn}a_{ijk} =
a_{ijn}\,a_{ikn}\,a_{rsn}\,a_{ikn}^{-1}\,a_{ijn}^{-1}, \quad  jk = rs.
$$
The other conjugation relations are found in the a similar way.

\begin{figure}
\vskip 1in
\caption{Conjugation in $P_n^2$} \label{movie}
\end{figure}

\s{The Reciprocity Law.}\label{recip}

In this section we define a move within the group $P_n^2$, which we
will call the {\it reciprocity law}. This will be used in \SC \ref{lift}
to lift certain relations from $P_{n-1}^2$ to $P_n^2$.

To understand the reciprocity law, it is helpful to understand
the analogous
move within the classical pure braid group. We may consider the generator
$a_{ij} \in P_n$ to be the braid whose $j$th string
passes around the $i$th string. However, by ``pulling the $j$th string
tight'' we can  also consider this braid to be the one whose $i$th string
passes around the $j$th string (see Figure \ref{pull}).
We label this new loop $\alpha_{ij}$. The reciprocity law for $P_n$ is simply
the statement
$a_{ij} = \alpha_{ij}$.

\begin{figure}
\vskip 1in
\caption{The reciprocity law for $P_n$} \label{pull}
\end{figure}

We describe the reciprocity law for $P_n^2$.
First we need to define a new loop $\alpha_{ijk}$  in $X_n^2$. Let
$F_j$ be the fiber defined in \SC \ref{fiber}.
Choose a point $q \in I(\R)$ on the line $L_{ik}(\R)$
within the disc of radius $\epsilon$ about the point $b_k$.
Let the line $L'(\R)$ denote the
real line joining $b_j$ and $q$.
Then we define the loop $\alpha_{ijk}$ to
be the loop in $L'$, based at $b_j$,
which goes around $q$.
To prove the reciprocity law for $P_n^2$ we will use the following key lemma.

\begin{figure}
\vskip 1.0in
\caption{Reciprocity} \label{keypic}
\end{figure}

\begin{lemma} \label{key}
The two loops $a_{i,k-1,k}$ and $\alpha_{i,k-1,k}$ are homotopic in
$X_n^2$ relative to the basepoint $b$, for $2 \le i < k \le n$ (see Figure
\ref{keypic}).
\end{lemma}

This proof is computational, and is given in \SC \ref{calc}.

\begin{figure}
\vskip 1.0in
\caption{The reciprocity law for $P_n^2$} \label{recproof}
\end{figure}

\begin{proposition} \label{rl}(The reciprocity law)
The loops $a_{ijk}$ and $\alpha_{ijk}$ are homotopic in $X_n^2$
relative to the basepoint $b \in B$, for $1 \le i < j < k \le n$ (see Figure
\ref{recproof}).
\end{proposition}

\begin{pf}
We find an explicit homotopy. Commence the homotopy by shrinking the loop
$a_{ijk}$. Let $\epsilon >0$
be a small real number. Let $L(\R)$ denote the line $L_k^{\infty}$ which was
used to define the loop $a_{ijk}$. Let
$p$ be a point on $\psi(\R)$ lying between the points $b_{k-1}$ and $b_k$.
Let $\tilde{L}(\R)$ be the line joining the points $b_j$ and $p$.
Let $v$ be a vector which orients the line $\tilde{L}(\R)$ in the directon from
$b_j$ to $p$.
Move the point $b_j$ up distance $\epsilon$ into $\tilde{L}$ in the direction
$iv$. Then move
$b_j$ along the line above $\tilde{L}(\R)$ until it reaches the point in
$\tilde{L}$
which is distance $\epsilon$ in the direction $iv$ above $p$. Finally move
$b_j$ back down to $p$.
During
the motion of $b_j$ the points $p_{ij} = L \cap L_{ij}$ will move within the
line
$L$. Shrink the loop $a_{ijk}$, so that it follows the motions of the $p_{ij}$.

By sliding points along $\psi(\R)$ whilst remaining in the base point set $B$,
if necessary, we
will now be in the same situation  as Lemma \ref{key} with $j=k$.
The shrunken version of $a_{ijk}$ is thus
homotopic to the loop $\alpha_{i,k-1,k}$.
By sliding points again, if necessary, we now move the point $b_j$ back
to its starting position along the same path which it originally took.
During this process
the loop $\alpha_{i,k-1,k}$ will get ``stretched out''.
The resulting loop is homotopic to $\alpha_{ijk}$.
\end{pf}

\s {Lifting relations}\label{lift}

In this \GS we show how to lift relations from the group $P_{n-1}^2$
to $P_n^2$ in order to complete the proof of  Theorem \ref{main theorem1}.
We also show how to lift relation (\ref{prod}) from $Q_{n-1}^2$ to $Q_n^2$.
Before we begin lifting relations, we describe the main idea that we shall use.

First, consider the classical pure braid groups. We may lift a relation
from $P_{n-1}$ to $P_n$, if the $n$th string does not obstruct the
homotopy describing this relation in $P_{n-1}$. Since the $n$th string does not
obstruct any of the homotopies describing the relations in $P_{n-1}$, they all
lift.

A similar idea applies in lifting relations from $P_{n-1}^2$
to $P_n^2$, although in this case we get some obstructions.
The point $b_n$ introduces the lines $L_{in}(\C)$ into $GF_{n}^2$.
If it is possible to describe a relation in $P_{n-1}^2$ by a homotopy within
$F_k$, $1 \le k \le n-1$, which does not intersect
any of the $L_{in}$, then we can lift this relation. This is because we have
the sequence\footnote{
See Section \ref{fiber} for the definition of $F_k$.}
$$
\pi_1(F_k) \to  P_n^2 \to P_{n-1}^2 \to 1.
$$
However,  if it is impossible to describe a relation in $P_{n-1}^2$
by a homotopy which does not intersect the  $L_{in}$,
then we have to use the reciprocity law to
lift the relation.

\st{Relations (\ref{comm1}) -- (\ref{comm5}).}

\begin{lemma}
For $1 \le k \le n-1$ the relations (\ref{comm1}) -- (\ref{comm5})
lift from $P_{n-1}^2$ to $P_n^2$.
\end{lemma}
\begin{pf}
For $1 \le k \le n-1$ the relations  (\ref{comm1}) -- (\ref{comm5}) hold in
$\pi_1(F_k)$.
\end{pf}

\st{Relation (\ref{conjs}).}

\begin{lemma}
For $1 \le k \le n-1$ the relation (\ref{conjs}) lifts from $P_{n-1}^2$ to
$P_n^2$.
\end{lemma}
\begin{pf}
Relation (\ref{conjs}) is described by a homotopy $H : [0,1] \times [0,1] \to
X_{n-1}^2$.
Let $R_l \subset L_l^{\infty}$ be equal to  $(\text{im}\, H)  \cap
L_l^{\infty}$.
Then the homotopy $H$ can be chosen so that the
lines $L_{in}$ within $F_k$ do not intersect the region $R_l$ for $l = j,k$.
This is because  the points $b_1, \dots , b_n$ satisfy the
the lexcigon condition. Hence $H$ lifts to $X_n^2$.
\end{pf}

\st{Relation (\ref{hopf}).}

This case is not the same as the previous two, since the
lines $L_{in}$ in $F_k$ obstruct the homotopy
describing relation (\ref{hopf}) within $P_{n-1}^2$. This is because the lines
$L_{in}$
always pass through the point $b_i$.
First note that for fixed $k$ and $j$ the relation\footnote{This relation
arises from the Hopf link of the point $b_j$.}
\begin{equation} \label{rec}
[a_{ijk},\dots ,a_{j-1,jk},\tilde{a}_{j,j+1,k},\dots
,\tilde{a}_{j,k-1,k},\alpha_{jk,k+1},\dots ,\alpha_{jk,n-1}] = 1
\end{equation}
holds in $\pi_1(F_k)$ (and thus in $P_n^2$).
To understand why this is the case see Figure \ref{why}.
Now using  the reciprocity law relation (\ref{rec}) can be rewritten as follows
$$
[a_{ijk},\dots ,a_{j-1,jk},\tilde{a}_{j,j+1,k},\dots
,\tilde{a}_{j,k-1,k},a_{jk,k+1},\dots ,a_{jk,n-1}] = 1.
$$
Thus relation (\ref{hopf}) holds $P_n^2$. Now we simply note that
$$
\ (p_n^2)_\ast([a_{ijk},\dots ,a_{j-1,jk},\tilde{a}_{j,j+1,k},\dots
,\tilde{a}_{j,k-1,k},a_{jk,k+1},\dots ,a_{jk,n-1}])=
$$
$$
 [a_{ijk},\dots ,a_{j-1,jk},\tilde{a}_{j,j+1,k},\dots
,\tilde{a}_{j,k-1,k},a_{jk,k+1},\dots ,a_{jk,n-2}],
$$
for $1 \le k \le n-1$. Hence we have lifted relation (\ref{hopf}).

\begin{figure}
\vskip 1in
\caption{How Hopf relations lift} \label{why}
\end{figure}

\st{Relation (\ref{prod}).} \label{prods}

We conclude this \GS by showing how to lift the product relations (\ref{prod})
from
$Q_{n-1}^2$ to  $Q_n^2$.
We can use the same method that we used to lift relation (\ref{hopf}).
Let $G_k$ denote the fiber of the projection, $Y_n^2 \to Y_{n-1}^2$, obtained
by forgetting the $k$th point. Then for each $k$ the relation (\ref{prod})
holds in
$\pi_1(G_k)$ as a consequence of the reciprocity law.
Thus relation (\ref{prod}) holds  in $Q_n^2$. Now note that,
$$
(p_n^2) _\ast (a_{12k}\,a_{13k}\dots a_{k-2,k-1,k}\,a_{1k,k+1}\dots
a_{k-1,k,k+1}\dots a_{1kn}\dots a_{k-1,kn}) =
$$
$$
a_{12k}\,a_{13k}\dots a_{k-2,k-1,k}\,a_{1k,k+1}\dots a_{k-1,k,k+1}\dots
a_{1k,n-1}\dots a_{k-1,k,n-1}.
$$
for $1 \le k \le n-1$. Hence we have lifted relation (\ref{prod}).

It is informative to understand how this relation lifts. First, let us
understand
the  analogous situation in
the pure braid group on $\P^1$. In Figure \ref{pureprod} we picture
how to lift the relation
\begin{equation} \label{Qlift}
a_{1,n-1}a_{2,n-1}\dots a_{n-2,n-1} =1,
\end{equation}
which holds in $Q_{n-1}$, to the relation
$$
a_{1n}a_{2n}\dots a_{n-1,n} = 1,
$$
which holds in $Q_n$. Note that the $n$th string will always obstruct any
homotopy
which describes relation (\ref{Qlift}) in the group $Q_{n-1}$.

\begin{figure}
\vskip 1.0in
\caption{Lifting a product from $Q_{n-1}$ to $Q_n$} \label{pureprod}
\end{figure}

We now show how to lift relation (\ref{prod}).
First, consider this relation in $Q_{n-1}^2$. This may be rewritten as the
product
\begin{equation} \label{x}
a_{12k}\dots a_{1,k,n-1}a_{23k}\dots a_{k-2,k-1,k}\alpha_{k,k+1,k+2}\dots
\alpha_{k,n-1,n-2}
\end{equation}
using the reciprocity law.
We see that a homotopy describing relation (\ref{prod})
can be chosen to lie in the line $L_k^{\infty}$ (Figure \ref{prodpic}).
However, the lines $L_{in}$  intersect
$L_k^{\infty}(\R)$, and so the product (\ref{x})  will be homotopic to a loop
which encircles
the points $L(\R) \cap L_{in}(\R)$, within $L_k^{\infty}$.
This is homotopic to the product
$$
\alpha_{k,n-1,n}^{-1}\dots \alpha_{k,k+1,n}^{-1}\alpha_{k-1,k,n}^{-1}\dots
\alpha_{1,k,n}^{-1}.
$$
The reciprocity law allows us to rewrite this expression in terms of the
$a_{ijn}$. Thus, we
have lifted relation (\ref{prod}).

\begin{figure}
\vskip 1.0in
\caption{ Lifting a product from $Q_{n-1}^2$ to $Q_n$} \label{prodpic}
\end{figure}

\s {Proof of Lemma \ref{key}}\label{calc}

We prove this lemma in the following way. In Section \ref{a}, we choose a
base point $x$ in the set $B \subset X_n^2$. We then define two loops $\xi_m$,
$\tilde{\xi}_m$, based at $x$, and
show that these loops are homotopic relative to $x$. In \SC \ref{b}, we
generalize
the result of Section \ref{a}. Finally, in Section \ref{c}, we show
how to use the results of Sections  \ref{a} and \ref{b} to prove Lemma
\ref{key}.

\ss{} \label{a}
We begin by choosing the base point $x$ in $B \subset X_n^2$. Note that
$ X_n^2 \subset (\C^2)^m $. The curve $\psi$ is
then given by the equation $\psi(t)= t^2$.
Let
$$
(x_m,y_m) = (-m/2n,\psi(-m/2n)) = (-m/2n,(m/2n)^2),\quad 1 \le m \le n-2.
$$
Choose $\epsilon \in \R$ to be a sufficiently small positive number.
Let $u = ((1-\epsilon),(1-\epsilon)^2)$ and
$v = ((1+\epsilon),(1+\epsilon)^2)$. Then, we define $x \in X_n^2$ to be the
point
$$
((x_1,y_1),\dots  ,(x_{n-2},y_{n-2}),u,v).
$$

We now define two loops $\xi_m$ and $\tilde{\xi}_m$ in $X_{n}^2$, and show that
they are homotopic relative to $x$.

Let $p_m$ be the $y$ co-ordinate of the intersection of the line
$x=(1+\epsilon)$ and
the line joining points $u$ and $(x_m,y_m)$. Then we have
$$
p_m = 2\epsilon((1-\epsilon)-(m/n)) +(1-\epsilon)^2.
$$
Let $q_m$ be the $y$ co-ordinate of the intersection of the line
$x=(1-\epsilon)$ and
the line joining points $v$ and $(x_m,y_m)$. Then we have
$$
q_m = -2\epsilon((1+\epsilon)-(m/n)) +(1+\epsilon)^2.
$$
Let
$$
r_m = (1+\epsilon)^2 - (p_m + p_{m+1})/2,
$$
and
$$
s_m = (q_m + q_{m+1})/2 - (1-\epsilon)^2.
$$
These will be the ``radii'' of the loops  $\xi_m$ and $\tilde{\xi}_m$
respectively (see Figure \ref{lloop}).

\begin{figure}
\vskip 1in
\caption{} \label{lloop}
\end{figure}

Define
$$
\xi_m(t) = ((1+\epsilon), (1+\epsilon)^2+r_m(e^{it}-1)/2) \text{\ \ for \ \ } 0
\le t \le 2\pi,
$$
and
$$
\tilde{\xi}_m(t) = ((1-\epsilon), (1-\epsilon)^2+s_m(1-e^{it})/2) \text{\ \ for
\ \ } 0 \le t \le 2\pi.
$$

A homotopy, $H_m(s,t)$, where $0 \le s \le 1$ and $0 \le t \le 2\pi$, between
the loops $\xi_m$ and $\tilde{\xi}_m$ is now given explicitly by
$$
H_m(s,t) = ((x_1,y_1),\dots
,(x_{n-2},y_{n-2})(u_1(s,t),u_2(s,t)),(v_1(s,t),v_2(s,t))),
$$
where
$$
(u_1(s,t),u_2(s,t)) =((1-\epsilon),(1-\epsilon)^2+(1-s)s_m(1-e^{it})/2),
$$
and
$$
(v_1(s,t),v_2(s,t)) = ((1+\epsilon),(1+\epsilon)^2+sr_m(e^{it}-1)/2).
$$

Note that $H_m(0,t) = \xi_m(t)$ and  $H_m(1,t) = \tilde{\xi}_m(t)$ for $0 \le t
\le 2\pi$.
Also, note that $H_m(s,0) = x$ and $H_m(s,2\pi) = x$ for $0 \le s \le 1$. We
are thus reduced to
showing that the homotopy lies in $X_n^2$.

Let $(x_1,y_1),\dots ,(x_n,y_n)$ be $n$ points in $\C^2$.
We will need to check whether three of these
points lie on a line. The
following condition will be convenient for our calculations. No
three of these points will lie on a line if and only if no $3 \times 3$ minor
of the matrix
$$
\left[
\begin{array}{cccc}
 1      &  1      & \cdots   &  1     \\
 x_1    &  x_2    & \cdots   &  x_n   \\
 y_1    &  y_2    & \cdots   &  y_n
\end{array}
\right]
$$
vanishes.
Hence, to prove that the loops $\xi_m$ and $\tilde{\xi}_m$ are homotopic, it
will be sufficient to show
that no $3 \times 3$ minor the matrix
$$
H(s,t) = \left[
\begin{array}{ccccc}
 1      &  \cdots & 1       &  1         &  1       \\
 x_1    &  \cdots & x_{n-2} &  u_1(s,t)  &  v_1(s,t) \\
 y_1    &  \cdots & y_{n-2} &  u_2(s,t)  &  v_2(s,t)
\end{array}
\right]
$$
vanishes for any $0 \le t \le 2\pi$ and $0 \le s \le 1$.

We split this calculation up into  three  cases. Let $C_i$ denote the $i$th
column of $H$, for $1 \le i \le n$. Since the points corresponding to
columns $C_1 , \dots , C_{n-2}$ lie on $\psi(\R)$ and are fixed for all $s$ and
$t$,
it suffices to check that the determinants of the following matrices
do  not vanish, for all $1 \le k < l \le n-2$:
\begin{equation} \label{d1}
  [C_k,C_l,C_{n-1}];
\end{equation}
\begin{equation} \label{d2}
  [C_k,C_l,C_{n}];
\end{equation}
\begin{equation} \label{d3}
  [C_k,C_{n-1},C_{n}].
\end{equation}

\sss {Determinant of (\ref{d1}).}
It suffices to check this only when  $t= \pi$, since
otherwise the determinant of matrix (\ref{d2}) has non-zero  complex part and
is thus non-zero.
Let $t$ be equal to $\pi$. Then matrix (\ref{d1}) is equal to
$$
\left[
\begin{array}{ccc}
 1          & 1         &  1            \\
 -(k/2n)    & -(l/2n)   &  (1-\epsilon) \\
 (k/2n)^2   & (l/2n)^2  & (1-\epsilon)^2+(1-s)s_m
\end{array}
\right]
$$
Thus we need only show that the determinant of this matrix does not vanish for
any $0 \le s \le 1$.
Expanding out the determinant of this matrix gives us the following expression:
$$
(k/2n - l/2n)( (kl/4n^2) +(1- \epsilon)^2 +(1-\epsilon)(l/2n+k/2n) - (1-s)s_m).
$$
Suppose that this expression is equal to zero for some $s$. Then, by
choosing $\epsilon$ sufficiently small we can force the inequality
$$
 s > (1+ 1/2n)^2.
$$
But $0 \le s \le 1$, and we are lead to a contradiction.

\sss {Determinant of (\ref{d2})}
It suffices to check this only when  $t= \pi$, since
otherwise  the determinant of matrix  (\ref{d2}) has non-zero  complex part and
is thus non-zero.
Let $t$ be equal to $\pi$. Then matrix (\ref{d2}) is equal to
$$
\left[
\begin{array}{ccc}
 1          & 1         &  1            \\
 -(k/2n)    & -(l/2n)   &  (1+\epsilon) \\
 (k/2n)^2   & (l/2n)^2  &  (1+\epsilon)^2+sr_m
\end{array}
\right]
$$
Thus we need only show that the determinant of this matrix does not vanish for
any $0 \le s \le 0$.
Expanding out the determinant of this matrix gives us the following expression:
$$
(k/2n-l/2n)( (kl/4n^2) +(1+ \epsilon)^2 +(1+\epsilon)(l/2n+k/2n) -sr_m).
$$
Suppose that this expression is equal to zero for some $s$. Then, by
choosing $\epsilon$ sufficiently small, we can force the inequality
$$
 s > (1+ 1/2n)^2.
$$
But $0 \le s \le 1$, and we are lead to a contradiction.

\sss{Determinant of (\ref{d3}).}

This case is more complicated since the imaginary part of the
determinant of matrix (\ref{d3})
may vanish. By simplifying the determinant, we obtain the following
expression for the imaginary part of the determinant of matrix (\ref{d3}):
\begin{equation} \label{stuff}
-\sin t \, \left( (s_m(1 + l/2n + \epsilon) + s((1+ l/2n - \epsilon)r_m
-(1+l/2n+ \epsilon)s_m))/2 \right).
\end{equation}
Note that $\sin t$ can only vanish when $t=0$
or when $t = \pi$. We explain what happens in the
case when $t = \pi$ later. Suppose that $t$ is not equal to $0$ or $\pi$ and
that expression (\ref{stuff}) vanishes.
In this case $s$ would be equal to
$$
-s_m(1+l/2n+\epsilon)/((1+l/2n-\epsilon)r_m -(1+l/2n+\epsilon)s_m).
$$
A computation shows that this quantity must be greater than one.
Hence, we are lead to a contradiction.

The final case we are left with is when $t = \pi$. In this case, by supposing
that
the determinant vanishes, we obtain:
\begin{equation} \label{e5}
sK = 1 - 2\epsilon(1+l/n-\epsilon)/s_m,
\end{equation}
where,
$$
K = (((1+l/2n-\epsilon)r_m)/((1+l/2n+\epsilon)s_m)-1).
$$
A computation shows that $K$
cannot be equal to zero, and hence we may divide equation (\ref{e5}) by $K$.
Again, we are left with $s$ equal to a quantity which, by allowing $\epsilon$
to be sufficiently small, may be shown to be greater than one.
Thus, the image of the map $H_k$ lies in $X_n^2$ and the loops $\xi_m$ and
$\tilde{\xi}_m$ are homotopic.

\ss{} \label{b}

We now generalize the result of Section \ref{a}. To prove Lemma \ref{key}, we
will need  to place points on the curve
$\psi(\R)$ to the right of the points $u$ and $v$, without obstructing the
homotopy
between the loops $\xi_m$ and $\tilde{\xi}_m$.
We do this by placing our extra points
on $\psi(\R)$ so that they
are ``far away'' from $u$ and $v$. In this way, we can
ensure that the lines introduced by adding in the new points do
not intersect the complexified lines $x = 1+ \epsilon$ and $x = 1 - \epsilon$
within
the circles of radius $r_m$ and $s_m$ respectively.
Now it is clear that the  lines introduced by adding the
new points will not obstruct the homotopy
between the loops $\xi_m$ and $\tilde{\xi}_m$.

\ss{} \label{c}

We now complete the proof of Lemma \ref{key}. Recall that
we are trying to show that the loops $a_{i,k-1,k}$ and $\alpha_{i,k-1,k}$
are homotopic relative to $b$. We can slide points up and down $\psi(\R)$ as
long as we
remain in the contractible set $B$ defined in \SC \ref{fiber}.
Slide the points $\{ b_{k+1},\dots b_n \}$ along the
curve $\psi(\R)$ until they are ``far to the right'' of the points $b_{k-1}$
and $b_{k}$. Now  slide points
$b_{k-1}$ and $b_{k}$ into positions $u$ and $v$ of \SC \ref{a}. Finally, slide
points $\{b_1,\dots,b_{k-2}\}$
so that we are in the situation of \SC \ref{b}. During the sliding process the
loops  $a_{i,k-1,k}$ and $\alpha_{i,k-1,k}$ will be deformed; we
denote the resulting loops by the same symbol.

The loop $\xi_m$ lies in the complexification of the line $x=1+\epsilon$. The
loop $a_{i,k-1,k}$ can
also  be homotoped into this line. Similarly, the loop $\alpha_{i,k-1,k}$
can be homotoped into the complexification of the line $x = 1-\epsilon$, in
which the loop $\tilde{\xi}_m$
lies. Hence, we are reduced to showing that certain loops are homotopic in
punctured complex lines. In this situation
we have
$$
a_{i,k-1,k} = \xi_i \xi_{i-1}^{-1},
$$
and
$$
\alpha_{i,k-1,k} = \tilde{\xi}_i  \tilde{\xi}_{i-1}^{-1}.
$$

We know from \SC \ref{b}  that the loops $\xi_m$ and $\tilde{\xi}_m$ are
homotopic for $1 \le m \le k-2$.
Thus the loop  $a_{i,k-1,k}$ is homotopic to the loop $\alpha_{i,k-1,k}$.
By sliding all points back to the basepoint $b$, whilst remaining in the set
$B$, we complete the proof of Lemma \ref{key}.

\end{document}